\documentclass[aps,pre,twocolumn,groupedaddress,showpacs,preprintnumbers,
amsmath,amssymb]{revtex4}
\newcommand{\beq}{\begin{equation}}
\newcommand{\eeq}{\end{equation}}
\newcommand{\bea}{\begin{eqnarray}}
\newcommand{\eea}{\end{eqnarray}}
\def\eref#1{(\ref{#1})}

\def\beq{\begin{equation}}
\def\eeq{\end{equation}}
\def\beqa{\begin{eqnarray}}
\def\eeqa{\end{eqnarray}}

\def\N{ {\cal N} }

\newcommand{\bv}{\boldsymbol v}

\usepackage{amsmath}
\usepackage{graphicx}
\title{}

\begin{document}

\title{Quantum Corrections to Fidelity Decay in Chaotic Systems  }

 \author{Boris Gutkin}
\affiliation{Fachbereich Physik, Universit\"at Duisburg-Essen, Lotharstra\ss e 1, D-47048 Duisburg, Germany}
 \author{Daniel Waltner}
\affiliation{Institut f\"ur Theoretische Physik, Universit\"at Regensburg, D-93040 Regensburg, Germany}
\author{Martha Guti\'errez}
\affiliation{Institut f\"ur Theoretische Physik, Universit\"at Regensburg, D-93040 Regensburg, Germany}
 \author{Jack Kuipers}
\affiliation{Institut f\"ur Theoretische Physik, Universit\"at Regensburg, D-93040 Regensburg, Germany}
 \author{Klaus Richter}
\affiliation{Institut f\"ur Theoretische Physik, Universit\"at Regensburg, D-93040 Regensburg, Germany}

\date{\today}

\begin{abstract}
By considering correlations between classical orbits we derive semiclassical
expressions for the decay of the quantum fidelity amplitude for classically chaotic quantum systems,
as well as for its squared modulus, the fidelity or Loschmidt echo.
Our semiclassical results for the fidelity amplitude agree with random matrix theory 
(RMT) and supersymmetry predictions in the universal Fermi golden rule regime. 
The calculated quantum corrections can be viewed as arising from a static 
random perturbation acting on nearly self-retracing interfering paths, and hence
will be suppressed for time-varying perturbations.
Moreover, using trajectory-based methods we show a relation, recently obtained in RMT,
between the fidelity amplitude and the cross-form factor for parametric level correlations.
Beyond RMT, we compute Ehrenfest-time effects on the fidelity amplitude.
Furthermore our semiclassical approach allows for a unified treatment of
the fidelity, both in the Fermi golden rule and Lyapunov regimes, 
demonstrating that quantum corrections are suppressed in the latter.
\end{abstract}
\maketitle

\section{Introduction}

Understanding the sensitivity of quantum time evolution in complex systems with respect to perturbations 
has evolved to a common subject of interest in the complementary fields of 
quantum information and quantum chaos. The concept of fidelity has become a central measure to quantify 
the stability of quantum dynamics upon perturbations. The fidelity $M(t)$ was introduced by Peres \cite{peres} as the
squared modulus of the fidelity amplitude, $m(t)$, the overlap integral of an initial state,
e.g.\ a wave packet, with the state obtained upon forward and backward propagation governed by two
Hamiltonians differing slightly by a perturbation. In the context of quantum chaos the fidelity is often
referred to as Loschmidt echo \cite{cit-MolPhysics}, picking up and generalizing concepts
from spin echo physics \cite{Abragam}. 

By definition the fidelity $M(t)$ equals unity at $t\!=\!0$ and usually decays further in time. 
Broad interest in fidelity decay had been initiated by the pionieering semiclassical work of Jalabert and
Pastawsky \cite{Jalabert} who found a perturbation-dependent decay for weaker perturbations
and discovered the intriguing Lyapnov regime for classically chaotic quantum systems under stronger perturbations.
Nowadays, three main decay regimes are distinguished, depending  
on the strength of the perturbation $\Sigma$: In the limit of a weak perturbation, i.e.\ if 
$\Sigma$ is smaller than the quantum mean level spacing $\Delta$, the fidelity decay is Gaussian
characterizing the {\em perturbative regime} \cite{jacq-1,toms,pros}. For perturbations of the order of or
larger than $\Delta$, the decay is predominantly exponential, $M(t) \sim {\rm e}^{-2\Gamma t}$,
with a decay constant $\Gamma$
obtained by Fermi's golden rule. The corresponding perturbation range is hence called the
{\em Fermi golden rule (FGR) regime} \cite{Jalabert,jacq-1,toms}. For strong perturbations the decay is 
still exponential in time, but with a $\Sigma$-independent decay rate given by the Lyapunov 
exponent $\lambda$ of the classically chaotic counterpart of the quantum system 
characterizing the {\em Lyapunov regime} \cite{Jalabert,Cucc}. 
For a review of the extensive literature on fidelity decay see Refs.~\cite{gorin,Petit}, 
including a discussion of further decay mechanisms.

Fidelity decay has been calculated within two main theoretical frameworks, namely random matrix
theory (RMT) and supersymmetrical approaches on one side, and semiclassical theory on the
other. Within RMT, the decay of the averaged fidelity amplitude was computed in the perturbative 
and FGR regime within
linear response \cite{gorin-2004} for weak perturbations. Later, using supersymmetric
techniques and going beyond linear response, these results were extended to stronger perturbations for the different universality classes \cite{Sto-2004,Sto-2005}. 
For the GUE case, the result for the fidelity amplitude in the universal
FGR regime takes the particularly compact form
\begin{equation}
 m_{\rm GUE} (\tau; \gamma)=
  \left\{\begin{array}{ll}
  -\frac{\partial }{\partial \gamma} \left(\frac{e^{-\gamma}}{\gamma\tau}
    \sinh(\gamma\tau)\right) & \mbox{ for } \tau<1 \, , \\
   -\frac{\partial }{\partial \gamma}\left(\frac{e^{-\gamma\tau}}{\gamma\tau}
    \sinh(\gamma)\right) & \mbox{ for } \tau>1 \, , \\
   \end{array} \right. 
    \label{RMTfidelity}
 \end{equation}
with $\gamma = \Gamma t$, $\tau=t/T_{\rm H}$ and $T_{\rm H}=2\pi\hbar\Delta$ is the  Heisenberg time.
In particular, these supersymmetrical calculations revealed a fidelity recovery at the Heisenberg time.
Recently, this has been interpreted by establishing relations between the fidelity amplitude 
decay and parametric level correlations \cite{Koh}, allowing fidelity decay phenomena to be considered
from a completely different perspective. There it was found that in the FGR regime 
the Fourier transform of the parametric level correlator, the cross-form factor $K(\tau;\gamma)$, 
and the RMT fidelity amplitude $m(\tau;\gamma)$ are closely related to each other through
\begin{equation}
\frac{\partial K(\tau;\gamma)}{\partial
\gamma}= - \frac{2\tau}{\beta} m(\tau;\gamma) \, ,
\label{relationship}
\end{equation}
for the different GOE, GUE and GSE RMT ensembles labeled by the index $\beta\!=\!1,2$ and 4.

Despite this recent progress, the averaged fidelity $M(t)$ itself, the
Loschmidt echo, has not been obtained using supersymmetric techniques. Within an RMT framework,
$M(t)$ was calculated within linear response which -- upon heuristically exponentiating
this result -- gives a fair approximation for the transition region between the perturbative
 and FGR regime \cite{gorin-2004}.
Both, RMT and supersymmetry approaches are limited in principle to the perturbative and universal (FGR) regime, 
governed by a single, system-independent perturbation parameter $\Gamma$, and cannot reveal information 
about the individual system, such as the Lyapunov decay for strong perturbations.
Hence, the Lyapunov regime is not amenable to RMT approaches.

The complementary situation appears from the viewpoint of semiclassical theory \cite{Jalabert,Cucc,Van03,Wen05}: 
While the existing semiclassical tools have lead to the discovery of the Lyapunov regime, 
$M(t) \sim {\rm e}^{-\lambda t}$  \cite{Jalabert},
uncovering an appealing connection between classical and quantum chaotic
dynamics, semiclassics has so far only been able to predict the leading exponential
decay in the FGR regime \cite{Jalabert}, but could not account for additional quantum corrections
arising from an expansion, 
in orders of $t/T_{\rm H}$, of the supersymmetry results such as Eq.~(\ref{RMTfidelity}).
The absence of such quantum interference contributions in the semiclassical approach can be
traced back to the so-called diagonal approximation used in the treatment of the fidelity amplitude
$m(t)$, that is a pairing of the same trajectories in a semiclassical path integral representation of 
the propagators for the perturbed and unperturbed system. 

In this paper we will go beyond this approximation by evaluating contributions from correlated 
trajectory pairs built from different orbits. Classical correlations between (periodic) orbits
were shown to be the key to understanding RMT predictions for spectral statistics \cite{Sie}
and to deduce universal spectral properties within a semiclassical theory.
This technique has been considerably further developed and applied to calculate various spectral 
\cite{Mul1,ref:Brouwer06B,Heu1,Kuip,Nag,Waltner09}, scattering and transport properties
\cite{Rich02,ref:Adagideli03,Heu,ref:Brouwer06,ref:Jacquod06,Ber}
of classically chaotic quantum systems \cite{WalRic09}. Recently, this method has been extended to semiclassically 
compute the quantum survival probability and photo-fragmentation processes of open chaotic systems
\cite{Wal,Gut}, as well as to establish a semiclassical version of the continuity equation \cite{Kui}.
Classical correlations encoded in the semiclassical diagrams considered in the works 
\cite{Wal,Gut,Kui} are of special relevance for ensuring unitarity in problems involving semiclassical 
propagation along open trajectories inside a system and will prove particularly important for
the fidelity decay.

We will show in Sec.~\ref{fidelity-amplitude} for the fidelity amplitude in the FGR regime 
for times below $T_{\rm H}$ that such subtle classical correlations also provide the 
semiclassical key to the aforementioned quantum corrections to the exponential decay of
$m(t)$, arising from an expansion of Eq.~(\ref{RMTfidelity}) in powers of $t/T_{\rm H}$. 
It is moreover of interest to explore the implications of the Ward identities
leading to Eq.~(\ref{relationship}) on the level of the semiclassical theory.
In Sec.~\ref{identities} we derive this relation by invoking recursion relations for 
terms containing the correlations between classical orbits.

Beside the formal derivation of Eq.~(\ref{RMTfidelity}), semiclassics also sheds light
on the underlying interference mechanism leading, e.g., to weak-localization-type
corrections to the exponential fidelity amplitude decay, encoded in Eq.~(\ref{RMTfidelity}). 
This arises from the fact that the relevant dynamics is organized along orbits with so-called 
encounter stretches, orbit segments which are traversed (time-shifted) twice in nearly opposite
directions. On its way back and forth the particle experiences practically the
same disorder perturbation potential. This gives rise to an enhanced single-particle
damping compared to dynamics along a generic orbit exploring independent
disorder regions, and hence finally leads to an enhanced fidelity decay.
This has interesting physical implications which we will discuss below:
(i) such interference effects will depend on the Ehrenfest time
where the correlation length of the disorder potential appears as a new relevant
length scale as will be discussed in Sec.~\ref{ehrenfest};
(ii) The above outlined investigation refers to a static (random) perturbation
as most of the works on fidelity decay. However, deviations are expected for  
time-varying perturbations, as e.g.\ caused by phonons. 
Our approach allows for incorporating such effects. We will show in Sec.~\ref{time-dep} 
how a decrease in the correlation time, i.e.\ faster time variations, gives rise to 
a stronger suppression of the quantum interference terms and thereby in turn to an {\em increase} of the 
overall fidelity amplitude.

In Sec.~\ref{fidelity} we address the fidelity $M(t)$, which is usually easier to measure.
We compute the corresponding quantum corrections for $M(t)$ and show that 
they are suppressed in the Lyapunov regime.
Our semiclassical approach hence adequately describes the FGR and Lyapunov fidelity regimes in 
a unified way.

\section{Fidelity amplitude}
\label{fidelity-amplitude}
\subsection{Semiclassical approach}

The fidelity amplitude $m(t)$ is defined as the overlap 
\begin{equation}
\label{eq-1}
m\left(t \right)=\int d\textbf{r}\, \Psi^* \left(\textbf{r},t \right)\Phi\left(\textbf{r},t \right), 
\end{equation}
between  states $\Psi \left(\textbf{r},t \right)$ and $\Phi\left(\textbf{r},t \right)$ obtained by propagating a common
initial state $\Psi_0$ with two slightly different Hamiltonians: $H$ and $H'=H \! + \! \Sigma$. The integration  in Eq.~(\ref{eq-1}) is performed over the available configuration
space, which in the case of  billiards is  the area enclosed by the boundaries. For  further semiclassical analysis of $m(t)$
it is convenient to represent  $\Psi \left(\textbf{r},t \right)$ with the help of the quantum propagator 
$K_H\left(\mathbf{r},\mathbf{r}',t\right)$ for the Hamiltonian $H$:
\begin{equation}
\label{eq-0.5}
\Psi \left(\textbf{r},t \right)= \int d\mathbf{r}' K_H\left(\mathbf{r},\mathbf{r}',t\right)\Psi_0 \left(\mathbf{r}' \right).
\end{equation} 
The same representation holds for $\Phi\left(\textbf{r},t \right)$, with  $K_H$ replaced by the  propagator $K_{H'}$ for the perturbed Hamiltonian $H'$.
The next step is to use the semiclassical Van Vleck formula \cite{Gutz} for the propagators $K_H$ and  $K_{H'}$. For two-dimensional systems one has
\begin{equation}
\label{eq0}
 K_{\rm sc}\left(\textbf{r},\textbf{r}',t \right)=\!\left(\frac{1}{2\pi i\hbar} \right)
\! \sum_{\tilde{\gamma}\left(\textbf{r}'\rightarrow\textbf{r},t\right)}\! C_{\tilde\gamma}^{1/2}
 \exp\left(\frac{i}{\hbar}S_{\tilde\gamma}\left(\textbf{r},\textbf{r}',t \right) \right),
\end{equation}
where $S_{\tilde\gamma}\left(\textbf{r},\textbf{r}',t \right)$ is the classical action of the trajectory
$\tilde\gamma$ running from $\textbf{r}'$ to $\textbf{r}$ in time $t$, and 
$C_{\tilde\gamma}=\left|\det\left(\partial^2 S_{\tilde\gamma}/\partial\textbf{r}' 
\partial\textbf{r}\right) \right|\exp\left({-\frac{i\pi}{2}\mu_{\tilde\gamma}}\right)$ 
is the Van Vleck determinant including the Maslov index $\mu_{\tilde\gamma}$.

We will  assume that the perturbation $\Sigma$ is classically
small such that only the actions, i.e.\ the phases, are affected while the classical trajectories $\tilde\gamma$  remain
unchanged. Under this assumption, 
after inserting Eqs.~(\ref{eq0}) and (\ref{eq-0.5}) into  Eq.~(\ref{eq-1}) 
we  obtain the following semiclassical approximation \cite{Jalabert} for $m(t)$:
\begin{eqnarray}
\label{eq0.5}
m_{\rm sc}(t)&=&\left(\frac{1}{2\pi\hbar} \right)^2 \int d  \textbf{r}d{\textbf r}' d{\textbf
r}''\Psi_0^*\left({\textbf r}'\right) \Psi_0\left({\textbf r}''\right)\\&&\!\!\times \!\!\!\!\!\!
\sum_{\tilde\gamma\left(\textbf{r}'\rightarrow\textbf{r},t\right),\atop\tilde\gamma'\left(\textbf{r}''\rightarrow\textbf{r},t\right)}\!\!\!\!\!\!\!
C_{\tilde\gamma}^{1/2}C_{\tilde\gamma'}^{*1/2}  \exp \left( {\frac{i}{\hbar} \left(
S_{\tilde\gamma}-S_{\tilde\gamma'}+\Delta S_{\tilde\gamma}\right) }\right), \nonumber
\end{eqnarray}
where $\Delta S_{\tilde{\gamma}}$ stands for  the change in the action along the trajectory $\tilde{\gamma}$ due to the perturbation $\Sigma$. 
Note that Eq.~(\ref{eq0.5}) involves a double sum over trajectories $\tilde\gamma$ and $\tilde\gamma'$  of the unperturbed Hamiltonian $H$ only. 
Due to the rapidly oscillating phase factor containing the action  differences 
$S_{\tilde\gamma}-S_{\tilde\gamma'}$ most of the contributions will cancel out except for semiclassically small 
action differences originating from  pairs of trajectories which  are  close to each other in the
configuration space. We thus can use a linear approximation in order to relate the actions $S_{\tilde\gamma}$, $S_{\tilde\gamma'}$ along   the trajectories  $\tilde\gamma$, $\tilde\gamma'$ to the actions $S_{\gamma}$, $S_{\gamma'}$ along the nearby trajectories $\gamma$, $\gamma'$ connecting the midpoint,
$\mathbf{r}_0=\left(\mathbf{r}'+\mathbf{r}''\right)/2$ with $\mathbf{r}$.
 This yields 
\begin{eqnarray}
\label{eq1}
m_{\rm sc}(t)&=&\left(\frac{1}{2\pi\hbar} \right)^2\!\! \int \!\!d  \textbf{r}d{\textbf r}_0 d{\textbf
q}\Psi_0^*\left({\textbf r}_0+ \frac{{\textbf q}}{2}\right) \Psi_0\left({\textbf r}_0- \frac{{\textbf q}}{2}\right)  \nonumber\\&&\times \sum_{\gamma\left({\textbf r}_0\rightarrow{\textbf r},t\right),\atop\gamma'\left({\textbf r}_0\rightarrow{\textbf r},t\right)} C_\gamma^{1/2} C_{\gamma'}^{*1/2}\exp\left( \frac{i}{\hbar} \left(  S_\gamma-S_{\gamma'}\right)\right)\nonumber\\ &&\times\exp\left[\frac{i}{\hbar}\left(-\frac{{\textbf q}}{2}\left({\textbf p}_0^\gamma+{\textbf p}_0^{\gamma'} \right)+\Delta S_\gamma \right)\right],
\end{eqnarray}
where ${\textbf q}=\left({\textbf r}'-{\textbf r}''\right)$, and ${\textbf p}_0^{\gamma}$ and 
${\textbf p}_0^{\gamma'}$ are the initial momenta of the trajectories $\gamma$ and $\gamma'$, 
originating from the expansion of the actions around ${\textbf r}_0$.
%
%

\subsection{Treatment of the perturbation}

Further evaluation of $m_{\rm sc}\left(t\right)$ requires us to consider the effect of the perturbation
on the phase difference  $\Delta S_\gamma$ in Eq.~(\ref{eq1}).  It can be expressed as 
\begin{equation}
\label{eqa}
\Delta S_\gamma=\int_0^t dt' L_\gamma^\Sigma\left(t'\right)
\end{equation} 
where  $L_\gamma^\Sigma\left(t'\right)$ is the difference between  the kinetic and potential energy of the perturbation $\Sigma$.
In the case of a perturbation  potential $\Sigma$, 
Eq.~(\ref{eqa}) simplifies to
\begin{equation}
\label{eqb}
\Delta S_\gamma(t)=-\int_0^t dt'
\Sigma\left[\mathbf{q}_\gamma\left(t'\right),t'\right] .
\end{equation}  
For fully chaotic systems the $\Delta S_\gamma(t)/t$ are distributed as  Gaussian random variables when $t$ is sufficiently large  (compared to all classical time scales), and     the variance is given by $t\int_{-\infty}^{\infty}d\tau C(\tau)$, where $C(\tau)\equiv\left\langle
L_\gamma^\Sigma \left(\tau\right) L_\gamma^\Sigma\left(0\right) \right\rangle$ is   time correlation  of
the perturbation. Assuming that the mean value of $L_\gamma^\Sigma$ is zero this implies 

\begin{eqnarray}
\label{eqe}
\left\langle \exp\left(\frac{i}{\hbar}\Delta S_\gamma\right)\right\rangle&=& 
\exp\left(-\frac{\left\langle\Delta S_\gamma^2 \right\rangle}{2\hbar^2}\right) \nonumber \\
%
%
%
& = & \exp\left(-\frac{ t}{2\tilde{t}}\right) \, ,
\label{eqh}
\end{eqnarray}
where
\begin{equation}
\label{eqi}
\frac{1}{\tilde{t}}=\frac{1}{\hbar^2}\int_{-\infty}^{\infty}d\tau C\left(\mathbf{\tau}\right).  
\end{equation}

Note that the average in Eq.~(\ref{eqe}) should be understood as the average in the phase space over 
trajectories $\gamma$ with different initial conditions. 
Alternatively, one can fix $\gamma$ and consider an average over an
ensemble of different perturbations of  $L_\gamma^\Sigma$.
One possible  realization of such  an ensemble is provided by a quenched static disorder potential
\begin{equation}
\label{eqc}   
\Sigma\left(\mathbf{r}\right)=\sum_{\alpha=1}^{N_i}
\frac{u_\alpha}{2\pi\xi^2}\exp\left[-\frac{1}{2\xi^2}\left(\mathbf{r}-\mathbf{R}_\alpha\right)^2\right] \, .
\end{equation}
It is given by $N_i$ random impurities in a cavity of an area $A$ with the Gaussian profile characterized by a finite correlation length $\xi$, 
see \cite{Jalabert,Ullmo}.
The independent impurities can be assumed to be uniformly distributed at positions $\mathbf{R}_\alpha$ with the 
densities $n_i=N_i/A$ and  strengths $u_\alpha$ obeying $\left\langle u_\alpha u_\beta \right\rangle=u^2\delta_{\alpha\beta}$.

%

The action differences $\Delta S_\gamma$ accumulated by segments of $\gamma$ separated by distances larger  than $\xi$ can be regarded as uncorrelated. Consequently the stochastic accumulation of  $\Delta S_\gamma$ along $\gamma$ can  be described by a random process, resulting in a Gaussian distribution of the action difference $\Delta S_\gamma$.  This 
yields the previous result (\ref{eqe}) with the noticeable difference that the average here is over the ensemble of the disorder potentials \cite{Jalabert,Ullmo}.
For the potential in Eq.~(\ref{eqc}) the decay time $\tilde{t}$ in Eq.~(\ref{eqh}) can be calculated explicitly. 
In this case the correlation function is given by 
\begin{eqnarray}
\label{eqj}   
C_\Sigma\left(\left|\mathbf{r}-\mathbf{r}'\right|\right)&=&\left\langle
\Sigma\left(\mathbf{r}(t)\right)\Sigma\left(\mathbf{r}(t')\right)\right\rangle\nonumber\\&=&
\frac{u^2n_i}{4\pi\xi^2}\exp\left[-\frac{1}{4\xi^2}\left(\mathbf{r}-\mathbf{r}'\right)^2\right] \, , 
\end{eqnarray}
and it depends only on the difference between $r=r(t)$ and  $r'=r(t')$. Using then ergodicity of the 
classical flow and substituting time averages in Eq.~(\ref{eqi}) by the integration over the cavity 
domain we obtain for the decay time the relation \cite{Ullmo}
\begin{equation}
\label{decay-time-xi}
 \tilde{t}_\xi =  \sqrt{2} \pi \ k \xi \ \tilde{t}_\delta  \, .
\end{equation}
Here, $k\!=\! mv_0/ \hbar \!=\! 2\pi / \lambda_B$, with $v_0$ the particle velocity and
$\lambda_B$ the de Broglie wave length, and 
$ \tilde{t}_\delta = \hbar^3 / (n_i u^2 m)$ the quantum elastic scattering time for the white-noise case of
$\delta$-scatterers. Note that for $k \xi \gg 1$, $\tilde{t}_\xi$ coincides with the elastic scattering
time obtained quantum mechanically within the first Born approximation for the disorder potential
(\ref{eqc}). In the limit  $k \xi \le 1$, where the semiclassical treatment of disorder effects in terms of 
unperturbed trajectories is no longer valid, Eq.~(\ref{eqe}) can still be used but with $\tilde{t}_\xi$ 
replaced by $\tilde{t}_\delta$.

\subsection{Evaluation within diagonal approximation}

In order to evaluate the double sum over paths in Eq.\ (\ref{eq1}) let us first consider only
the diagonal contribution by pairing each path with itself, 
i.e.\ arising from the terms $\gamma=\gamma'$.
After inserting Eq.~(\ref{eqh}) into Eq.\ (\ref{eq1})  we obtain  \cite{Jalabert}: 
\begin{eqnarray}
\label{eql}
m_d\left(t\right)&=&\left(\frac{1}{2\pi\hbar} \right)^2\!\! \int \!\!d \textbf{r}d{\textbf r}_0 d{\textbf
q}\,\Psi_0^*\left({\textbf r}_0+ \frac{{\textbf q}}{2}\right) \Psi_0\left({\textbf r}_0- \frac{{\textbf
q}}{2}\right)\nonumber\\&& \sum_{\gamma\left({\textbf r}_0\rightarrow{\textbf r},t\right)}
\left|C_\gamma\right|  \exp \left( {-\frac{i}{\hbar} {\textbf q}{\textbf p}_0^\gamma  } \right) \exp \left(
-\frac{t}{2\tilde{t}} \right) \, .
\end{eqnarray}
The prefactor $\left|C_\gamma\right|$ can be regarded as a Jacobian when transforming  the integral over 
the final position $\mathbf{r}$ into an integral over the initial momentum 
${\mathbf p}_0\equiv{\textbf p}_0^\gamma$ of each trajectory. We then obtain 
\begin{equation}
\label{eqm}
m_d\left(t\right)=\left\langle \exp \left( -\frac{t}{2\tilde{t}} \right)\right\rangle_{{\mathbf r}_0,{\mathbf p}_0}
\, .
\end{equation}
Here $\left\langle \ldots\right\rangle_{{\mathbf r}_0,{\mathbf p}_0}$ indicates the phase space average~\cite{Gut},
\begin{equation}
\label{eqn}
\left\langle F\right\rangle_{{\mathbf r}_0,{\mathbf p}_0}= \frac{1}{(2 \pi \hbar)^2}
\int d{\mathbf r}_0 d{\mathbf p}_0\, F\left({\mathbf r}_0,{\mathbf p}_0\right)\rho_W\left({\mathbf r}_0,{\mathbf p}_0\right), 
\end{equation} 
where
\begin{eqnarray}
\label{eqo}
\rho_W\left({\mathbf r}_0,{\mathbf p}_0\right)&=&\int d{\mathbf q}\,\Psi_0^*\left({\mathbf r}_0+\frac{{\mathbf q}}{2}\right)\Psi_0\left({\mathbf r}_0-\frac{{\mathbf q}}{2}\right)\nonumber \\&& \times\exp\left(-\frac{i}{\hbar}{\mathbf q}{\mathbf p}_0\right)
\end{eqnarray}
denotes the Wigner function of the initial wave packet $\Psi_0$.
Assuming that $\Psi_0$ has a small energy dispersion around a mean energy $E_0$, Eq.~(\ref{eqm}) 
can simply be replaced by
\begin{equation}
\label{eqp}
m_d\left(t \right)=\exp\left(-\Gamma t  \right) \, ,
\end{equation}
with the decay rate $\Gamma\equiv 1/ (2\tilde{t}\left(E_0\right)) $. 
This represents the exponential FGR decay in the universal regime \cite{Jalabert,jacq-1,toms}.
As our expressions for the fidelity amplitude depend on $\Gamma$, we will also use in the following
the notation $m\left(t;\Gamma \right)$ instead of $m(t)$.

\begin{figure}
\begin{center}
\includegraphics[width=0.482\textwidth]
{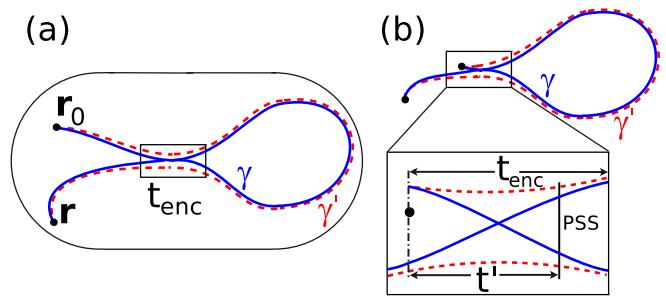}
\caption{Sketch of pairs of correlated classical trajectories $\gamma$ and $\gamma'$. 
In (a) the encounter region (rectangular box) connects a loop with two long legs beginning at $\bf r_0$ and ending at
$\bf r$. In contrast the paths begin or end inside the encounter region (`one-leg-loops') in 
(b). The zoom into the encounter region 
in (b) illustrates the position of the Poincar\'e surface of section (PSS) used and the definition of the 
encounter time.}\label{cap:fg4}
\end{center}
\end{figure}

\subsection{Loop corrections}

We now consider contributions to Eq.~(\ref{eq1}) from pairs of  
different trajectories following each other closely in configuration space, shown as full and dashed
line in Fig.\ \ref{cap:fg4}. The structure of such pairs can be characterized by long links, where  two 
trajectories  almost coincide, and  by  encounter regions (the boxes in Fig.\ \ref{cap:fg4}) 
where the segments of two trajectories are connected in a different way. Correlated trajectory pairs of
this type were first introduced in the context of spectral statistics involving periodic orbits 
by Sieber and Richter \cite{Sie} showing how universal RMT predictions can be obtained based 
only on semiclassical methods. We already mentioned important extensions of this approach in the
introduction. Within the phase space framework developed in Refs.~\cite{Mul1} we will closely follow 
the lines of Refs.~\cite{Wal,Gut}, where this formalism was employed in a semiclassical approach to
decay and fragmentation processes.
The main difference to the present case consists in the fact that the quantum  correction to 
the classical exponential survival probability in Refs.~\cite{Wal,Gut} (due to trajectory pairs 
containing a common encounter region) has to be replaced by a different correction 
to Eq.~(\ref{eqm}) which has to be computed.
One has to account for the fact that the disorder induced phase difference is modified for correlated 
trajectory pairs, because the same perturbation acts twice within the encounter region.
Although an expression for such a phase difference is known in a general context \cite{Nag,Kuip} 
we first illustrate how it can be calculated in the case of an orbit with the simplest possible encounter 
shown in Fig.\ \ref{cap:fg4}a. 
To this end we split the orbit into parts consisting of  two encounter-stretches and  three links,
\begin{eqnarray}
\label{eq3}
&&\hspace*{-1.5em}\left\langle \exp \left[ \frac{i}{\hbar}\Delta S_\gamma\right]  \right\rangle\nonumber\\&=&
\left\langle \exp \left[ \frac{i}{\hbar}\left( \sum_{i=1}^3\Delta S_{l(i)}+\sum_{i=1}^2\Delta S_{e(i)}\right) \right]  \right\rangle \nonumber\\ &=&
\left\langle \exp \left[ \frac{i}{\hbar}\left( \sum_{i=1}^3\Delta S_{l(i)}+2\Delta S_{e(1)}\right) \right]
\right\rangle \, .
\end{eqnarray}
In the last line we assume that, in the semiclassical limit and for a fixed disorder correlation
length $\xi$, the nearly parallel encounter stretches in the box are so close to each other (at a distance smaller
than $\xi$) that they experience the same perturbation. Considering a Gaussian phase distribution 
for each orbit segment (encounter stretch and three links) and  neglecting further correlations between the 
different segments we finally apply Eq.~(\ref{eqh}) to every part of the orbit individually and obtain
\begin{eqnarray}
\label{eq3.5}
&&\hspace*{-1.5em}\left\langle \exp \left( \frac{i}{\hbar}\Delta S_\gamma\right)  \right\rangle\nonumber\\&=&
\exp \left[ -\left( \dfrac{\sum_{i=1}^3\left\langle \Delta S_{l(i)}^2\right\rangle }{2\hbar^2} +4\dfrac{\left\langle\Delta S_{e(1)}^2 \right\rangle }{2\hbar^2} \right)\right] \nonumber\\ &=&
\exp\left[-\Gamma\left(t+2t_{\rm enc}\right)  \right]  \, .
\end{eqnarray}
%
Here $t_{\rm enc}$ denotes the length of the encounter region. 

Apart from the basic diagrams in 
Fig.\ \ref{cap:fg4}a we will also consider trajectory pairs that differ in an arbitrary number of encounters 
with an arbitrary number of stretches involved. For quantifying the encounter structure of an orbit, 
we introduce notation as in Refs.~\cite{Heu,Mul1}. 
We define a vector ${\bv}=\left(v_2,v_3,\ldots\right)$, 
where the component $v_l$ gives the number of $l$-encounters, i.e.\ the number of times 
where $l$ stretches of an orbit come close to each other. We denote by $V=\sum_{l=2}^\infty v_l$ the overall 
number of encounters and by $L=\sum_{l=2}^\infty lv_l$ the overall number of encounter stretches of an orbit. 
Repeating now the steps in Eqs.~(\ref{eq3},\,\ref{eq3.5}) for the general case of multiple encounters
one finds \cite{Nag,Kuip}
\begin{equation}
\label{eq14}
\left\langle \exp \left( \frac{i}{\hbar}\Delta S_\gamma\right)  \right\rangle=
\exp\left[-\Gamma\left(t+\sum_{\alpha=1}^V\left(l_\alpha^2-l_\alpha\right)t_{\rm enc}^\alpha\right)  \right], 
\end{equation} 
where $l_\alpha$ denotes the number of stretches of the encounter $\alpha$. We note that the structure of the
correction to the survival probability in Refs.~\cite{Wal,Gut} is slightly different,
\begin{equation}
\label{eq15}
\exp\left[-\Gamma_{\rm d} \left(t-\sum_{\alpha=1}^V\left(l_\alpha-1\right)t_{\rm enc}^\alpha\right)  \right],
\end{equation} 
where $\Gamma_{\rm d}$, the inverse dwell time $1/\tau_d$, takes the role of $\Gamma$.
Using Eq.~(\ref{eq3.5}), we can now evaluate  in Eq.~(\ref{eq1}) the off-diagonal contributions to the 
sums over trajectory pairs differing at encounters in the same way as in Refs.~\cite{Wal,Gut,Kui}.
Here we sketch some details of these calculations. In the region around 
each $l$-encounter we consider a Poincar\'e surface of section and measure the differences $s_j$ and 
$u_j$ between the piercing points of one of the trajectories along the stable and unstable manifolds, respectively. 
In these coordinates the duration of the encounter $\alpha$ can be expressed as \cite{Mul1}
\begin{equation}
\label{eq15.4}
 t_{\rm enc}^\alpha=\frac{1}{\lambda}\ln\left(\frac{c^2}{\max_j \left|u_{j,\alpha}\right| \max_j
 \left|s_{j,\alpha}\right|}\right) \, ,
\end{equation}
and the action difference between  two trajectories 
is given by 
\begin{equation}
\label{eq15.5}
S_\gamma - S_{\gamma'} =\sum_{\alpha=1}^V\sum_{j=1}^{l_\alpha-1} s_{j,\alpha}u_{j,\alpha} \, .
\end{equation}
Similarly to Refs.~\cite{Gut,Kui} we have to distinguish three different cases when the density of encounters is considered:  
(A) encounters inside the trajectory (Fig.\ \ref{cap:fg4}a), 
(B) encounters at the beginning or the end of the trajectory (Fig.\ \ref{cap:fg4}b) 
and (C) different encounters at the beginning \textit{and} at the end of the trajectory.

For  trajectories of type (A) and duration $t$ whose encounters are characterized by $\bv$, 
the corresponding weight is given by 
\begin{equation}
\label{eq16}
w_{t,A}\left(s,u \right)=N\left(\bv\right)\dfrac{\left[t-\sum_{\alpha=1}^V l_\alpha t_{\rm enc}^\alpha
\right]^L}{L!\:\Omega^{L-V} \prod_{\alpha=1}^V t_{\rm enc}^\alpha} \, .
\end{equation}
Here, $\Omega$ is the volume of the available phase space, 
and $N\left(\bv\right)$ is the number of trajectory structures, i.e.\ the number of topological 
realizations of orbits with the vector $\bv$, see Ref.~\cite{Mul1}. (Note that, in contrast to Refs.~\cite{Gut,Kui},
we include $N\left(\bv\right)$ into the weight function.) 
For a trajectory with  the beginning or end point inside of  an encounter, the weight function can be 
conveniently expressed as 
\begin{equation}
\label{eq17}
w_{t,B}\left(s,u \right)=\dfrac{2N\left(\bv\right)\sum_{\alpha=1}^V l_\alpha t_{\rm
enc}^\alpha\left[t-\sum_{\alpha=1}^V l_\alpha t_{\rm enc}^\alpha  \right]^{L-1}}{L!\:\Omega^{L-V}
\prod_{\alpha=1}^V t_{\rm enc}^\alpha} \, .
\end{equation}
Finally, for trajectories with  both beginning and end point inside of two \textit{different} encounters we have 
\begin{equation}
\label{eq18}
w_{t,C}\left(s,u
\right)=\!\!\sum_{\alpha,\beta=1\atop\alpha\neq\beta}^V \! \dfrac{\N_{\alpha\beta}\left(\bv\right)t_{\rm enc}^\alpha
t_{\rm enc}^\beta\left[t-\sum_{\alpha=1}^V l_\alpha t_{\rm enc}^\alpha
\right]^{L-2}}{\left(L-2\right)!\:\Omega^{L-V} \prod_{\alpha=1}^V t_{\rm enc}^\alpha} \, .
\end{equation}
Here $\N_{\alpha\beta}\left(\bv\right)$ is defined as the number of possible ways to cut links between the 
encounters $\alpha$ and $\beta$ divided by $L$ in all periodic orbit structures described by the vector $\bv$, 
see Ref.~\cite{Gut,Kui}.

We can now present general expressions for the loop contributions to the fidelity amplitude resulting 
from the three weight functions in Eqs.~(\ref{eq16}-\ref{eq18}). In view of Eqs.~(\ref{eq14}), (\ref{eq15.5}) 
and Eq.~(\ref{eq16}), we obtain in case (A) the following contribution to the fidelity amplitude from  
trajectories of time $t$ with encounters structure characterized by $\bv$:
\begin{eqnarray}
\label{eq22}
m_{A}\left(t;\Gamma,\bv\right)&=&N\left(\bv\right)\int_{-c}^{c}d\mathbf{s}d\mathbf{u}\dfrac{\left[t-\sum_{\alpha=1}^V
l_\alpha t_{\rm enc}^\alpha  \right]^L}{L!\Omega^{L-V} \prod_{\alpha=1}^V t_{\rm enc}^\alpha}{\rm
e}^{\frac{i}{\hbar}\mathbf{s}\mathbf{u}}\nonumber\\
&\times& \exp\left[-\Gamma\left(t+\sum_{\alpha=1}^V\left(l_\alpha^2-l_\alpha\right)t_{\rm
enc}^\alpha\right)\right] \, .
\end{eqnarray}
Applying then the rule \cite{Mul1} that, after expansion  of Eq.~(\ref{eq22}) in $t_{\rm enc}^\alpha$, 
the only non-vanishing contributions come from the terms independent of $t_{\rm enc}^\alpha$, we get 
\begin{eqnarray}
\label{eq23}
m_{A}\left(t;\Gamma,\bv\right)&=&\left(\frac{1}{T_{\rm H}\Gamma}\right)^{L-V}N\left(\bv\right)\prod_{l=2}^\infty\left(-\frac{l}{t}\frac{\partial}{\partial \Gamma}-l^2\right)^{v_l}\nonumber\\ &&\times\frac{\left(\Gamma t\right)^L\exp\left(-\Gamma t\right)}{L!},
\end{eqnarray}
where we used $T_{\rm H}=2\pi\hbar \Delta $ and the definition of components $v_l$.
In a similar way we can derive from Eq.~(\ref{eq17}) the contribution  for the case (B):
\begin{eqnarray}
\label{eq24}
m_{B}\left(t;\Gamma,\bv\right)&\!=\!&2\left(\frac{1}{T_{\rm H}\Gamma}\right)^{L-V}N\left(\bv\right)\sum_{l_1=2}^\infty\frac{l_1v_{l_1}}{L}\nonumber\\&&\times\left(-\frac{l_1}{t}\frac{\partial}{\partial
\Gamma}-l_1^2\right)^{v_{l_1}-1}   \\ &&\times\prod_{l=2\atop l\neq
l_1}^\infty\!\!\left(-\frac{l}{t}\frac{\partial}{\partial
\Gamma}-l^2\right)^{v_l}\!\!\frac{\left(t\Gamma\right)^{L-1}\exp\left(-\Gamma t\right)}{\left(L-1\right)!} \,
, \nonumber
\end{eqnarray} 
with an $l_1$-encounter at the beginning or end of the trajectory. Finally we obtain from Eq.~(\ref{eq18})
for the case (C)
\begin{eqnarray}
\label{eq25}
m_{C}\left(t;\Gamma,\bv\right)&\!=\!&\left(\frac{1}{T_{\rm H}\Gamma}\right)^{L-V}\sum_{l_1,l_2=2}
^\infty\left(-\frac{l_1}{t}\frac{\partial}{\partial \Gamma}-l_1^2\right)^{v_{l_1}-1}\nonumber\\&&
\times\N_{l_1l_2}\left(\bv\right)\left(-\frac{l_2}{t}\frac{\partial}{\partial
\Gamma}-l_2^2\right)^{v_{l_2}-1}\\&&\times\prod_{l=2\atop l\neq l_1,
l_2}^\infty\left(-\frac{l}{t}\frac{\partial}{\partial \Gamma}-l^2\right)^{v_l}\frac{t^{L-2}\exp\left(-\Gamma
t\right)}{\left(L-2\right)!} \, , \nonumber
\end{eqnarray}  
with an $l_1$-encounter and an $l_2$-encounter at the beginning and at the end of the trajectory, respectively. 

The entire  contribution from orbit pairs to $m\left(t;\Gamma\right)$ is obtained 
from Eqs.~(\ref{eq23}-\ref{eq25}) by summing over all possible vectors $\bv$ and finally adding up
 three contributions together with the diagonal term (\ref{eqp}):
\begin{eqnarray}
\label{fidsm}
m_{\rm sm}\left(t;\Gamma\right) &=& m_d\left(t;\Gamma\right) + \sum_{\bv}\Big[ m_{A}\left(t;\Gamma,\bv\right) \nonumber\\ 
&& + m_{B}\left(t;\Gamma,\bv\right) + m_{C}\left(t;\Gamma,\bv\right)\Big] \, . 
\end{eqnarray} 

We now compare the contributions to the fidelity amplitude resulting from our semiclassical expressions
(\ref{eq23}-\ref{eq25}) with the RMT results \cite{Sto-2004,Sto-2005}. 

For systems with time-reversal symmetry the leading order correction is due to
 the 2-encounter diagrams in Fig.\ \ref{cap:fg4}. From these diagrams we obtain as one main result
\begin{equation}
\label{eq19}
m^{q_1}_{GOE}(t;\Gamma)=\left(-\frac{\Gamma t^2}{T_{\rm H}} \right) \exp\left(-\Gamma t \right) \, .
\end{equation}
This leading quantum interference correction leads to a reduction of the fidelity amplitude.
Equation~(\ref{eq19}) indicates that quantum deviations of the fidelity from pure exponential 
decay become relevant at time scales 
$t \sim (\tilde{t} T_{\rm H})^{1/2}$, with decay time $\tilde{t}\! =\! 1/2\Gamma$, i.e.\ at 
times which depend on the perturbation strength and 
can be much shorter than the Heisenberg time $T_{\rm H}$. Thus the quantum corrections can occur at timescales before saturation sets in \cite{peres,Cucc,FALS}.

\begin{figure}
\begin{center}
(a)~\includegraphics[height=2.0cm]{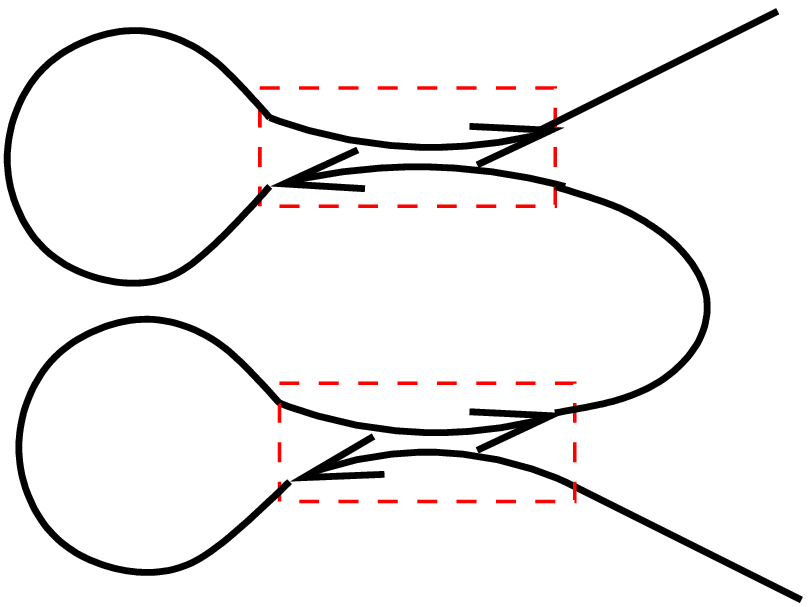}
\hskip 0.0cm
\includegraphics[height=2.2cm]{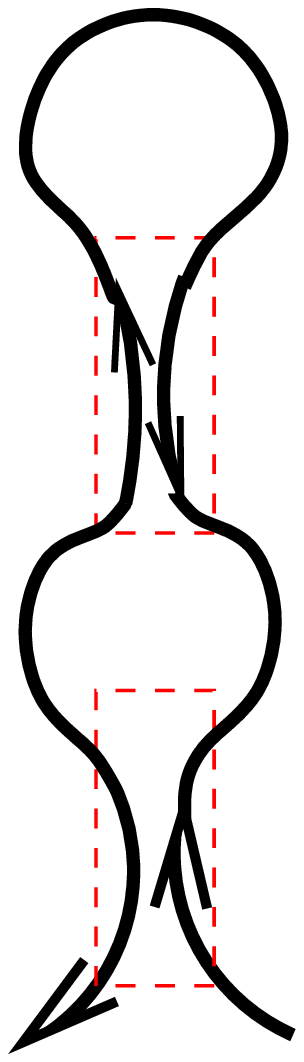}~(b)
\hskip 0.0cm
\includegraphics[height=2.0cm]{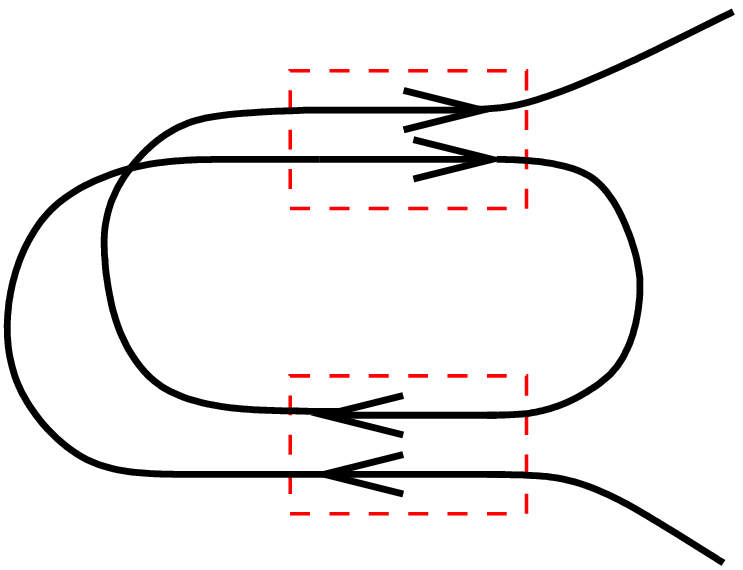}~(c)
\hskip 0.2cm
  (d)~\includegraphics[height=2.0cm]{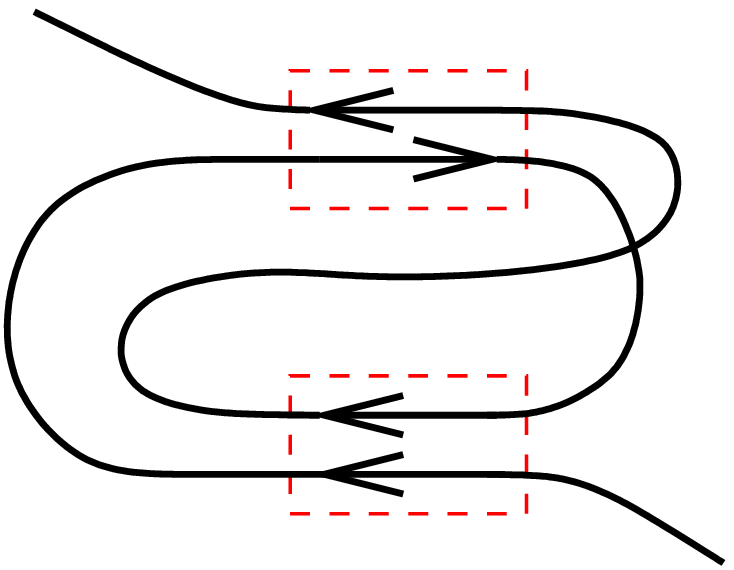}
\hskip 0.0cm
\includegraphics[height=2.0cm]{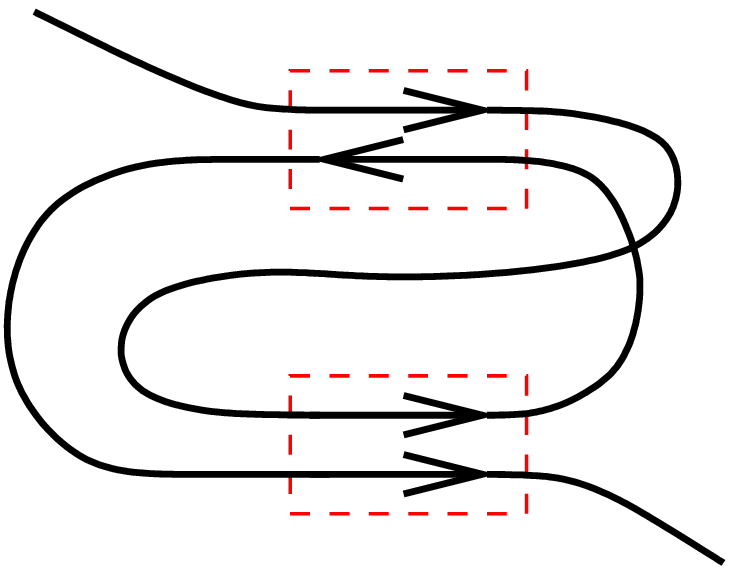}~(e)
\end{center}
\caption{Diagrams with two 2-encounters. In the unitary case only the diagram (c) contributes to the fidelity amplitude. }\label{cap2a}
\end{figure}

\begin{figure}
\begin{center}
(a)~\includegraphics[height=1.6cm]{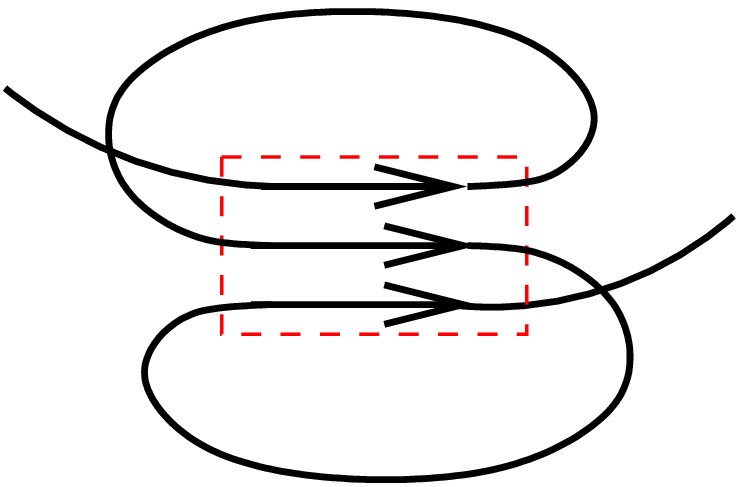}
\hskip 0.0cm
\includegraphics[height=1.5cm]{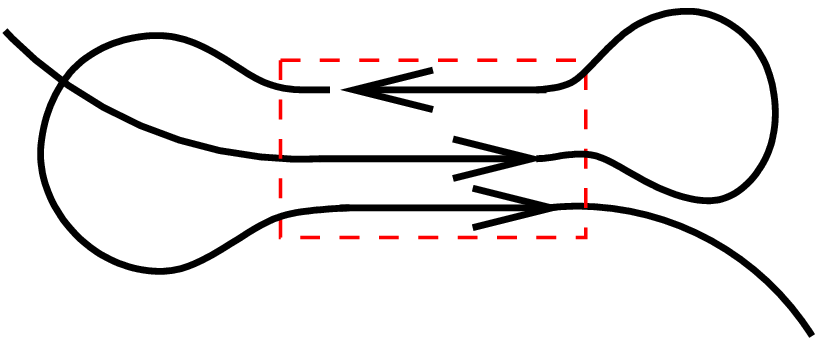}~(b)
\hskip 0.2cm
(c)~\includegraphics[height=1.5cm]{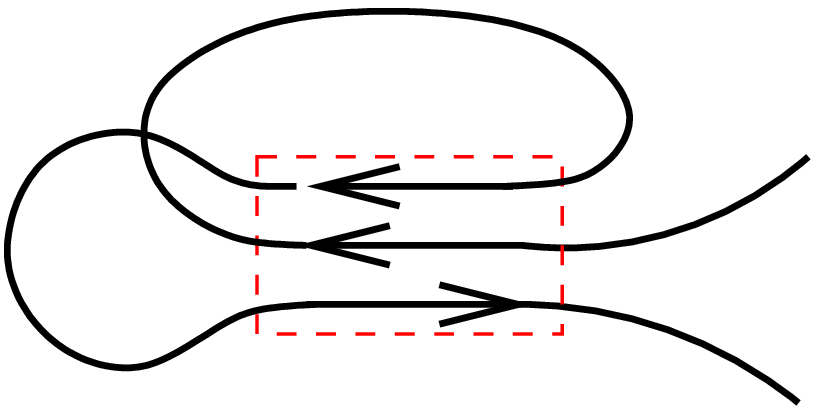}
\hskip 0.0cm
\includegraphics[height=1.5cm]{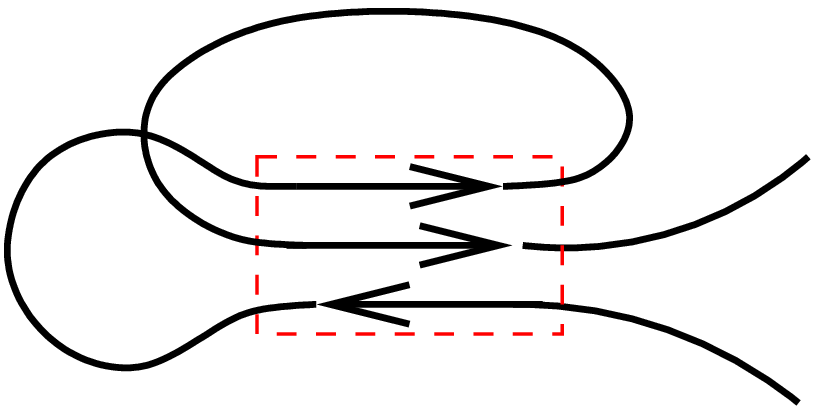}~(d)
\end{center}
\caption{Diagrams with one 3-encounter. In the unitary case only the diagram (a) contributes to the fidelity amplitude. }\label{cap2b}
\end{figure}

The next-to-leading order contributions come from the diagrams of higher order with either two 2-encounters
(as in Fig.\ \ref{cap2a}) or with one 3-encounter  (as in Fig.\ \ref{cap2b}) \cite{Gut}. 
All together they give:
\begin{equation}
\label{eq21}
m^{q_2}_{GOE}(t;\Gamma)=\left(-2\frac{\Gamma t^3}{3 T_{\rm H}^2 }+\frac{5\Gamma^2 t^4}{6T_{\rm H}^2} \right)\exp\left(-\Gamma t \right).
\end{equation}
In the unitary case the leading order  correction originates only from the diagrams shown in Fig.\ \ref{cap2a}c and Fig.\ \ref{cap2b}a,   yielding 
%
%
\begin{equation}
\label{eq20}
m^q_{GUE}(t;\Gamma)=\left(-\frac{\Gamma t^3}{3 T_{\rm H}^2 }+\frac{\Gamma^2 t^4}{6T_{\rm H}^2} \right)\exp\left(-\Gamma t \right) \, .
\end{equation}
%
The results (\ref{eq19}-\ref{eq20}) are consistent with the ones obtained from RMT-calculations, as can be
seen by expanding in $t/T_{\rm H}$ the RMT expressions  for $m(t;\Gamma)$, $t/T_{\rm H}<1$ in Refs.~\cite{Sto-2004,Sto-2005} and comparing the 
leading-order terms.

It is instructive to analyze these results and the underlying semiclassical assumptions for the
case of the disorder potential (\ref{eqc}) with correlation length $\xi$. To generate random (Gaussian)
fluctuations, $\xi$ should be smaller than the system size $\cal L$. Furthermore, in the semiclassical
limit $\xi \gg \lambda_B$. During the derivation of the above equations we implicitly assumed 
that the distance between encounter stretches, of the order of $\sqrt{{\cal L}\lambda_B}$, is smaller 
than $\xi$ in order to have correlated disorder along the stretches and thereby loop interference
corrections. This implies that our approach is valid for parameters obeying 
\begin{equation}
\label{eq-xi-L}
\sqrt{{\cal L}\lambda_B} < \xi < \cal L \, .
\end{equation}
For white noise disorder, $\xi \rightarrow 0$, the semiclassical approach would predict
vanishing interference corrections, since then the disorder potential along the two encounter stretches
is no longer correlated. This would imply clear deviations from the universal RMT result for
the white noise case. In that case the present semiclassical picture of orbits which remain unaffected by
the disorder (up to the phases) has to be replaced by a semiclassical approach based on 
paths scattered at impurities. We believe that presumably Hikami boxes take the role of the encounter regions
finally establishing universality for the white noise case.

\section{Identities for the fidelity amplitude}
\label{identities}

In this section we address two important properties of  the semiclassical quantum fidelity amplitude $m_{\rm sc}$ and prove them within the trajectory-based 
formalism for times $t<T_{\rm H}$. The first one: $m_{\rm sc}(t;\Gamma\!=\!0) \equiv 1$, is the direct  consequence of the unitarity of the semiclassical evolution. The second one is the connection  (\ref{relationship})  between the fidelity amplitude and the parametric spectral form factor. As we will show below, from the semiclassical point of view both properties can be attributed to the existence of certain  
recursion relations satisfied by  $N\left(\bv\right)$ and $\N_{kl}\left(\bv\right)$.

The fact that we can prove to all orders both the unitarity of semiclassical evolution and Eq.~(\ref{relationship})
shows the consistency of our semiclassical approach for times below $T_{\rm H}$.

\subsection{Unitarity}

In the case of vanishing perturbation ($\Gamma=0$) the fidelity amplitude $m(t)$ should be equal to one by the unitarity of quantum evolution.  In the following we will show that this property holds for the semiclassical form of $m(t)$ obtained in the previous section when $t<T_{H}$. 

As one can immediately see from Eq.~(\ref{eqp}), $m_{\rm sc}(t;\Gamma=0) =1$ within the diagonal approximation. 
Hence one has to show that all further semiclassical loop contributions vanish for $\Gamma=0$. 
To this end we will demonstrate that the off-diagonal terms
$m_A(t;0,\bv), m_B(t;0,\bv)$ and  $m_C(t;0,\bv)$ cancel each other.
According to Eqs.~(\ref{eq23},\ref{eq24},\ref{eq25}) they read:
\begin{eqnarray}
\label{eq26.3}
m_A\left(t;0,\bv\right)&=&N\left(\bv\right)\left(-1\right)^V\! \left(\frac{t}{T_{\rm H}}\right)^{L-V}\! \frac{\left(\prod_{\alpha=1}^V l_\alpha\right)}{\left(L-V\right)!} \, , \nonumber\\
&&\\
\label{eq26.4}
m_{B}\left(t;0,\bv\right)&=&2N\left(\bv\right)\left(-1\right)^{V-1}\frac{V}{L}\left(\frac{t}{T_{\rm H}}\right)^{L-V}\nonumber\\ &&\times\frac{\left(\prod_{\alpha=1}^V l_\alpha\right)}{\left(L-V\right)!} \, , \\
\label{eq26.5}
m_{C}\left(t;0,\bv\right)&=&\sum_{k,l=2}^\infty\frac{\N_{kl}\left(\bv\right)}{lk}\left(-1\right)^{V-2}\left(\frac{t}{T_{\rm H}}\right)^{L-V} \nonumber\\ && \times\frac{\left(\prod_{\alpha=1}^V l_\alpha\right)}{\left(L-V\right)!} \, .
\end{eqnarray}
Adding these contributions, one can see that the off-diagonal corrections disappear if for each $\bv$ the following condition is satisfied:
\begin{equation}
\label{eq26.6}
 N\left(\bv\right)-2\frac{V}{L}N\left(\bv\right) +\sum_{k,l=2}^\infty\frac{\N_{kl}\left(\bv\right)}{kl} =0 \, .
\end{equation}
Precisely the same unitarity condition was obtained in Ref.~\cite{Kui} for the continuity equation. 
As it was shown there, Eq.~(\ref{eq26.6}) can  be proven by relating $\N_{kl}\left(\bv\right)$ to $N\left(\bv\right)$ via the relationship 
\beq \label{twoencorbeqiveqn}
\N_{kl}(\bv)=\frac{(k+l-1)(v_{k+l-1}+1)}{\left(L-1\right)}N(\bv^{[k,l\to k+l-1]}) \, ,
\eeq
where $v_{k+l-1}$ is the $(k+l-1)$-th component of $\bv$, and $\bv^{[k,l\to k+l-1]}$ is the vector obtained by 
decreasing the components $v_k$ and $v_l$ by one and increasing the component $v_{k+l-1}$ by one. 
To obtain Eq.~\eref{twoencorbeqiveqn} one considers link  contraction such that a $k$-encounter and a $l$-encounter merge into a $(k+l-1)$-encounter. By looking at the number of possible ways to contract the link and to form a smaller 
periodic orbit structure, Eq.~\eref{twoencorbeqiveqn} can be deduced from the results in Refs.~\cite{Heu,Mul1}.

In view of Eq.~\eref{twoencorbeqiveqn}, Eq.~\eref{eq26.6} can be transformed into 
\begin{equation}
 \left( L-2V-\sum_{k,l}\bv_{k+l-1}\right)N\left(\bv\right) =0
\end{equation}
which can be proven after performing the double sums over $k$ and $l$ \cite{Kui}.

\subsection{Relation between the fidelity amplitude and the parametric spectral form factor} 

The  surprising  connection  (\ref{relationship}) between the quantum fidelity amplitude and parametric 
spectral correlations was derived in Ref.~\cite{Koh} within an RMT approach. The idea was to use a  certain 
invariance property of the integration measure for the ensemble of random matrices. A corresponding Ward 
identity then led to Eq.~(\ref{relationship}), which can equivalently be put into the form
 \begin{equation}
\label{eq27}
-\frac{T_{\rm H}\beta}{2t^2}\frac{\partial K\left(t; \Gamma\right)}{\partial\Gamma}=m\left(t; \Gamma\right)\, .
\end{equation}
Here we show how this relationship can be obtained in the framework of our semiclassical approach 
for systems with and without time reversal symmetry ($\beta=1,2$). 
To this end it is convenient to work with the Laplace transforms of the semiclassical expressions
for the form factor and fidelity.

In order to reveal a systematic structure for the contribution of each encounter stretch and each link, 
it is instructive to take the Laplace transform of Eq.~(\ref{fidsm})  
with respect to $\gamma\equiv\Gamma t$ while keeping  
$\eta\equiv t/\left(T_{\rm H}\gamma\right) = (T_{\rm H}\Gamma)^{-1}$  fixed:  
\begin{equation}
\label{eq26}
\tilde{F}\left(q,\eta\right)=\int_0^\infty d\gamma \,m_{\rm sm}\left(\eta T_{\rm H}\gamma;\left(\eta T_{\rm H}\right)^{-1}\right)\exp\left(-q\gamma\right).
\end{equation}
Inserting the expressions for $m_{\rm sm}\left(t;\Gamma\right)$ and performing the Laplace transformation by partial integration gives rise to 
the perturbative expression (in powers of $\eta$)
\begin{equation}
\label{eq26.1}
\tilde{F}\left(q,\eta\right)=\sum_{n=0}^\infty \tilde F_n\left(q\right)\eta^n,
\end{equation}
where the $n$-th term in this expansion originates from trajectory pairs with $L-V=n$. Explicitly, the terms 
$\tilde{F}_n\left(q\right)$ take the form 
\begin{eqnarray}
\label{eq26.2}
\tilde{F}_n\left(q\right)&=&\sum_{|\bv|=n}\dfrac{\prod_{l=2}^\infty\left(-lq-l^2\right)^{v_l}}{\left(q+1\right)^{L+1}}\Bigg[N\left(\bv\right)\nonumber\\&& \left.+2\sum_{l_1=2}^\infty N\left(\bv\right)\frac{l_1v_{l_1}}{L}\left(\frac{q+1}{-l_1q-l_1^2}\right)\right.\nonumber\\&& \left.+\sum_{l_1,l_2=2}^\infty\dfrac{\N_{l_1l_2}\left(\bv\right)\left(q+1\right)^2}{\left(-l_1q-l_1^2\right)\left(-l_2q-l_2^2\right)}\right],
\end{eqnarray}
where  the sum is over all diagrams with fixed $|\bv|=L-V$.
In Eq.~\eref{eq26.2} we recognize the following diagrammatic rule: every link contributes to
$\tilde{F}_n\left(\eta\right)$ a factor $\left(q+1\right)^{-1}$ and every $l$-encounter a factor 
$-l(q+l)$. 
%

Now we turn to the parametric spectral form factor for which we use 
the following semiclassical expression \cite{Nag,Kuip}:
\begin{eqnarray}
\label{eq28}
K\left(t; \Gamma\right)&=&\frac{2}{\beta}\frac{\Gamma t^2}{T_{\rm H}}\sum_{\bv} \left(\frac{1}{\Gamma T_{\rm H}}\right)^{L-V}\prod_{l=2}^\infty\left(-\frac{l}{t}\frac{\partial}{\partial \Gamma}-l^2\right)^{v_l}\nonumber\\&&\times N\left({\bv}\right)\frac{\left(\Gamma t\right)^{L-1}e^{-\Gamma t}}{L!} \, .
\end{eqnarray}

After taking the Laplace transform (\ref{eq26}) of the left hand side of Eq.~(\ref{eq27}) we obtain for the $n$-th term of the expansion (in powers of  $\eta$) 
\begin{equation}
\label{lhseqn1}
\tilde{F}'_n=\left(n+q\frac{\partial}{\partial q}\right)\sum_{|\bv|=n}\frac{\tilde{N}(\bv,q)}{L},
\end{equation}
%
%
where $\tilde{N}(\bv,q)$ is defined as
\beq\label{tilden}
\tilde{N}(\bv,q)=\frac{N(\bv)(-1)^{V}}{(q+1)^{L}}\prod_{l}\left(l(q+l)\right)^{v_{l}}.
\eeq
Upon performing  the derivatives in Eq.~\eref{lhseqn1}  we obtain

\beq \label{lhssimplifiedeqn} 
\tilde{F}'_n=\sum_{|\bv|=n}\frac{\tilde{N}(\bv,q)}{L}\left[\frac{L}{(q+1)}-V+\sum_{l}\frac{qv_{l}}{(q+l)}\right] \, .
\eeq
This must be compared with the  $\tilde{F}_n$ term, Eq.~(\ref{eq26.2}),
for the Laplace transform of the fidelity amplitude. 
Using Eq.~(\ref{tilden}) we can rewrite it as
\begin{eqnarray}
\label{eq30}
& &\tilde{F}_n= \sum_{|\bv|=n}\frac{\tilde{N}(\bv,q)}{L}\left[\frac{L}{(q+1)}-\sum_l\frac{2 v_{l}}{(q+l)}\right] \\ 
& & {} + (-1)^{V-2}\left(\sum_{k,l}\frac{\N_{kl}(\bv)}{k(q+k)l(q+l)}\right)\frac{\prod_{l}\left(l(q+l)\right)^{v_{l}}}{(q+1)^{L-1}} \, . \nonumber
\end{eqnarray}
Furthermore, we can simplify this by expressing the matrix elements $\N_{kl}\left(\bv\right)$ in terms of $\tilde{N}(\bv,q)$.
Using Eq.~\eref{twoencorbeqiveqn} and 
%
%
taking into account the additional $q$-dependent factors  we get
\beqa
\tilde{F}_n&=&\left(-1\right)^{V-2}\frac{\N_{kl}(\bv)}{k(q+k)l(q+l)}\frac{\prod_{l}\left(l(q+l)\right)^{v_{l}}}{(q+1)^{L-1}}\nonumber \\ 
 &=& -\frac{\tilde{N}(\bv^{[k,l\to k+l-1]},q)}{\left(L-1\right)}\frac{v_{k+l-1}^{[k,l\to k+l-1]}}{(q+k+l-1)}.
\eeqa
We can then rewrite the sum over the dummy vectors, $\bv'=\bv^{[k,l\to k+l-1]}$, as a sum over $\bv$ which gives for the second line in Eq.~\eref{eq30}:
\beq
-\sum_{|\bv|=n}\frac{\tilde{N}(\bv,q)}{L}\left(\sum_{k,l}\frac{v_{k+l-1}}{(q+(k+l-1))}\right).
\eeq
Using then the results of Ref.~\cite{Kui} and performing the sum over $k$ it can be further  simplified to
\beq
-\sum_{|\bv|=n}\frac{\tilde{N}(\bv,q)}{L}\left(\sum_{l}\frac{(l-2)v_{l}}{(q+l)}\right),
\eeq
leading to the following expression for $\tilde{F}_n$:
\beq \label{rhssimplifiedeqn}
\tilde{F}_n=\sum_{|\bv|=n}\frac{\tilde{N}(\bv,q)}{L}\left[\frac{L}{(q+1)}-\sum_{l}\frac{lv_{l}}{(q+l)}\right].
\eeq
The final step is to show that Eq.~(\ref{rhssimplifiedeqn}) coincides with  Eq.~(\ref{lhssimplifiedeqn}).  
This is indeed so, since the difference between the two expressions,

\beqa 
&&\sum_{|\bv|=n}\frac{\tilde{N}(\bv,q)}{L}\left[V-\sum_{l}\frac{qv_{l}}{(q+l)}-\sum_{l}\frac{lv_{l}}{(q+l)}\right]\nonumber\\ 
 &=&\sum_{|\bv|=n}\frac{\tilde{N}(\bv,q)}{L}\left[V-\sum_{l}v_{l}\right] \, ,
\eeqa
vanishes due to the fact that $V=\sum_{l}v_{l}$. This completes the prove of relationship (\ref{eq27})
with semiclassical means.  This result has the added bonus of showing that we recover the fidelity amplitude in Eq.~(\ref{RMTfidelity}) for $\tau<1$ for systems with broken time reversal symmetry as we know that the semiclassical and RMT parametric form factor exactly agree in that regime \cite{Nag,Kuip}. 

To show semiclassically that Eq.~(\ref{eq27}) also holds in the symplectic case ($\beta=4$) we consider spin-orbit interaction for spin $1/2$-particles. In this case the spectral form factor $K(\tau;\gamma)$ on the left hand side of Eq.~(\ref{eq27}) is modified to $-\frac{1}{2}K_{GOE}(-\tau/2;\gamma)$ \cite{Mul1} with $K_{GOE}(\tau;\gamma)$ being the form factor in the GOE-case. The right hand side yields for spin-orbit interaction the additional factor $\left(-1/2 \right)^{L-V}$ that was derived in \cite{Gut} from the results obtained in \cite{Bol}. A short calculation then shows that Eq.~(\ref{eq27}) also holds in the symplectic case. However, we note here that it is not true for general spin $s$-particles.

%


\section{Ehrenfest time dependence of the fidelity amplitude}
\label{ehrenfest}

The Ehrenfest time $\tau_E$ (see e.g.\ Ref.~\cite{ref:Chirikov}) is the timescale a minimal wave-packet 
needs to spread in the phase space of a chaotic system  to a size such that it can no longer be described by a single classical trajectory.  As was pointed out by Aleiner and Larkin 
\cite{ref:Aleiner96} in the context of transport problem,  $\tau_E$ is the minimal time required for quantum effects to appear. Because of this 
Ehrenfest-time effects  on  stationary transport processes have been  a subject of considerable  interest,  both theoretically 
\cite{ref:Aleiner96,ref:Adagideli03,ref:Brouwer06,ref:Jacquod06} and experimentally 
\cite{ref:Yevtushenko00}. Furthermore, signatures of the Ehrenfest time were also studied in the time 
domain \cite{ref:Brouwer06B,ref:Schomerus04,ref:SchoJach05,Wal,Gut}, 
where they are expected to be particularly  pronounced.  
In this section we investigate $\tau_E$-effects on the fidelity decay.
Namely, we will consider the $\tau_E$-dependence of 
the first quantum correction to the fidelity amplitude, Eq.~(\ref{eq19}), 
for systems with time reversal symmetry. Our treatment  will follow the 
lines of Refs.~\cite{Wal,Gut}.

\begin{figure}
\begin{center}
\includegraphics[width=0.482\textwidth,bb=0 0 720 453]
{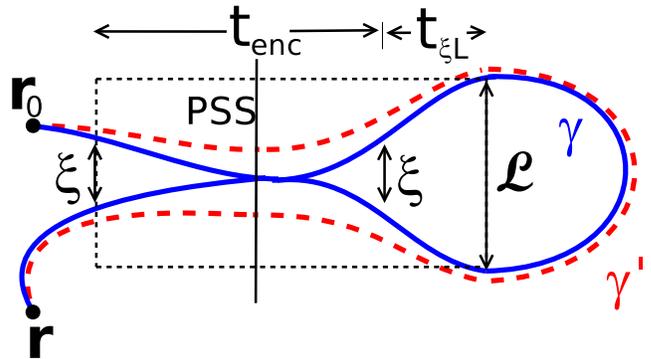}
\caption{Sketch of trajectories $\gamma$ and $\gamma'$ with a 2-encounter illustrating the length
restrictions in the case of finite Ehrenfest time.}\label{cap:fg5}
\end{center}
\end{figure}

Below we derive Ehrenfest time corrections to  $m_{GOE}^{q_1}$ coming from the trajectories of the type (A), 
see Fig.~\ref{cap:fg5}.
To this end we have to specify the classical constant $c$ 
appearing in Eq.~(\ref{eq15.4}).
Note that the different encounter stretches are subject to uncorrelated disorder when their spatial distance 
becomes larger than $\xi$. We thus require that inside of an encounter two segments of the trajectory are separated by distances less then $\xi$, see Fig.~\ref{cap:fg5}.  Accordingly, the encounter time is defined by 
\beq
t_{\rm enc}\equiv\frac{1}{\lambda}\ln\left(\frac{\xi^2\hbar}{\mathcal{L}\lambda_B\left|su
\right| } \right) \, ,
\label{t_enc}
\eeq
where $\cal L$ is the system size. The range of validity of our  approach is given by
relation (\ref{eq-xi-L}).

The densities (\ref{eq16}) and (\ref{eq17}) should be multiplied by a 
Heaviside function in time ensuring that a contribution exists only if the trajectory time $t$
is sufficiently long to enable a closed path. Specifically, on the right hand side 
of the encounter, depicted in Fig.\ \ref{cap:fg5}, the stretches 
should be separated by a distance of the order of ${\mathcal L}$ 
in order to close themselves and form a loop. 
On the 
left side, however, the encounter stretches have to be separated only by the distance $\xi$ (as we consider the case (A)).
This means the minimal time of the trajectory is $2t_{\rm enc}+2t_{\rm \xi L}$, where 
\beq
t_{\rm \xi L}=\lambda^{-1}\ln(\mathcal{L}/\xi)
\label{tau_Lx}
\eeq 
is the time it takes the stretches to be separated 
by the distance ${\mathcal L}$ when they are initially separated by the distance $\xi$. 

Accordingly, the weight function (\ref{eq16}) is slightly modified by introducing these minimal times and takes the form
\beq\label{w2llte}
w_{t,A}(u,s)=\frac{(t-2(t_{\rm enc}+t_{\rm \xi L}))^2}{2\Omega t_{\rm enc}}\theta(t-2t_{\rm
enc}-2t_{\rm \xi L}) \, .
\eeq
To account for the correction due to the action difference 
in Eq.~(\ref{eq14}) we again use $t_{\rm enc}$ as defined in Eq.~(\ref{t_enc}). Now we are in the
position to calculate the Ehrenfest time corrections 
to $m_{GOE}^{q_1}(t;\Gamma)$ in the case (A). As before, we will consider the Laplace transform (\ref{eq26}) of the fidelity amplitude. With the above corrections it reads
\bea
\widetilde{F}_{A}^{q_1}\left(\frac{\alpha}{\Gamma},\Gamma\right)=\int_{0}^{\infty} dt {\rm
e}^{-\left(\alpha+{\Gamma}\right)(t+2t_{\rm \xi L})}\frac{\Gamma t^2}{T_{\rm H}}I(\alpha,t) \, ,
\eea
where we shifted the integration range by $2t_{\rm enc}+2t_{\rm \xi L}$, and the integral
\beq
I(\alpha,t)=\frac{2}{\pi\hbar}\int_{0}^{c} du\int_{0}^{c} ds\frac{{\rm e}^{\frac{i}{\hbar}us}}{t_{\rm enc}} \exp\left(-2\alpha t_{\rm enc}-
4{\Gamma t_{\rm enc}}\right)
\eeq
 can be evaluated as in App.\ B of Ref.~\cite{Gut}. 
In a similar way the calculations can be performed for
trajectory pairs of type B with encounters at the beginning 
or at the end of the trajectory.

Summing up all the contributions and taking the inverse Laplace transform we  obtain the final result for the Ehrenfest time dependence of the entire,
leading quantum correction in the orthogonal case:
\bea
m_{\rm sc}^{q_1}(t;\Gamma)=-\frac{\Gamma}{T_{\rm H}}(t-2\tau_E^d)^2{\rm e}^{-\Gamma (t+2\tau_E^{\xi})}\theta(t-2\tau_E^d),
\label{m_ehrenfest}
\eea
where  $\tau_E^{\rm c}\!=\!\lambda^{-1}\!\ln({\mathcal L}/\lambda_B)$, $\tau_E^{\xi}\!=\!\lambda^{-1}\!\ln(\xi^2/({\mathcal L}\lambda_B))$ 
and $2\tau_E^d=\tau_E^{\rm c}+\tau_E^{\xi}$. 
Equation~(\ref{m_ehrenfest}) reveals
the role of finite Ehrenfest times corrections: they lead to
a time shift such that the quantum corrections set in later, i.e., for short times the system behaves ``classically''. 

Note also that corresponding calculations for the spectral parametric correlations give a 
time delay $2\tau_E^{\rm c}$ which is independent of $\xi$, since in that case only periodic 
trajectories are considered. This demonstrates that in general the connection (\ref{relationship}) 
between the fidelity amplitude and spectral parametric correlations breaks down when non-universal effects
such as the Ehrenfest time corrections are taken in account. 


\section{Fidelity amplitude for a time-dependent perturbation}
\label{time-dep}

Many physical situations require a generalization of fidelity decay to systems with a time-varying
perturbation. This is relevant if a subsystem evolves under the influence of a time-dependent external
environment or may arise if one models fidelity decay for a many-body system
in terms of single-particle dynamics exposed to an external fluctuating potential mimicking
the mutual interactions. Fidelity decay for time-dependent perturbations has been addressed in
Refs.~\cite{Jacetal01,Benenti02} in numerical studies of periodically kicked systems which can be 
regarded as time-dependent, in Ref.~\cite{Zurek} in the context of
decoherence and in Ref.~\cite{Cuc06}, which represents a direct extension of the semiclassical
approach of Refs.~\cite{Jalabert,Cucc} to a spatially and time-varying random potential.

In Ref.~\cite{Cuc06}, both a finite disorder correlation length $\xi$  (and associated time $\xi/v_0$ with
$v_0$ the particle velocity) as well as a finite correlation time $\tau_0$ characterizing
temporal fluctuations was introduced.
It was shown semiclassically on the level of the diagonal approximation, how the FGR decay is
governed by a decay rate $\Gamma$ into which both time scales generally enter and which is 
predominantly determined by the shorther of the two times, $\xi/v_0$ or $\tau_0$, if they 
strongly differ.

In this section we will use the perturbation model of Ref.~\cite{Cuc06} and illustrate the effect of 
a spatially and time-dependent perturbation for the representative case of the first quantum 
correction, $m^{q_1}_{GOE}(t;\Gamma)$, see Eq.~(\ref{eq19}) for the static case. 
Since this interference contribution is based on the mechanism that the same (static) perturbation 
exists along the two encounter stretches traversed at different times we can compute how finite 
correlation times $\tau_0$ will reduce this effect.

We will consider the interesting case where $\xi/v_0 \ll \tau_0$ so that $\Gamma$ depends only 
on the spatial fluctuations and not on $\tau_0$.
(For pure temporal fluctuations or $\xi/v_0 \gg \tau_0$, a change in $\tau_0$ will alter the
exponential decay rate $\Gamma$ and thereby mask the effect of $\tau_0$ on quantum fidelity 
contributions.)
We assume that the time dependence and the spatial dependence of the perturbation
can be separated, i.e.\ 
\begin{equation}
\Delta S_\gamma=\int_0^t dt'\,V\left( \textbf{q}(t')\right)W(t') \, ,
\end{equation}
which allows for further analytical treatment.
We first consider the averaged phase difference for this perturbation
\begin{eqnarray}
\label{eqab}
&&\hspace*{-2.1em}\left\langle\exp\left(\frac{i}{\hbar}\Delta S_\gamma\right) \right\rangle\nonumber\\&=& \exp\left(-\frac{2}{\hbar^2}\left\langle\int_0^{t_{\rm enc}} dtdt' V\left( \textbf{q}(t)\right)W(t)\right.\right.\nonumber\\&&\times V\left( \textbf{q}(t_l+2t_{\rm enc}-t')\right)W(t_l+2t_{\rm enc}-t') \Bigg\rangle \Bigg)\nonumber\\ &&\times \exp\left(-\Gamma t\right)
\end{eqnarray}
with a loop of length $t_l$ connecting the two encounter stretches.
The contribution in the last line of this equation is the overall exponential decay, not taking into account the correlated way the perturbation acts during the encounter.  The additional effects of this correlation are included in the exponential in the second and third line in the equation above, and yielded a contribution proportional to ${\rm e}^{-2\Gamma t_{\rm enc}}$ for an explicitly time-independent perturbation.  However we now have to analyze this contribution in more detail: First we again use that the two stretches during the encounter are very close together implying $\textbf{q}(t_l+2t_{\rm enc}-t')\approx\textbf{q}(t')$. 
Furthermore, we assume in this section, as in Ref.~\cite{Cuc06}, a Gaussian form of the spatial and the 
time-dependent perturbation
\begin{eqnarray}
&&\left\langle V\left( \textbf{q}(t)\right)W(t)V\left( \textbf{q}(t_l\!+\!2t_{\rm enc}\!-\!t')\right)
 W(t_l\!+\!2t_{\rm enc}\!-\!t')\right\rangle \nonumber\\ 
&&= \frac{\Gamma v_0\hbar^2}{\xi\sqrt{\pi}}\exp\left(-\frac{\left|\textbf{q}(t)-\textbf{q}(t')
\right|^2}{\xi^2}\right) \nonumber\\
&&\quad \times\exp\left(-\frac{\left(t_l+2t_{\rm enc}-t'-t \right)^2}{\tau_0^2} \right). 
\end{eqnarray}
The two time integrals in Eq.~(\ref{eqab}) are transformed into one integral with respect to $\tau=t-t'$ and one
with respect to $\tau'=(t+t')/2$. The integral with respect to $\tau$ is performed from $-\infty$ to $\infty$
assuming that the correlation length of the spatial part, $V\left( \textbf{q}(t)\right)$, 
is much shorter than the one of the time dependent part $W(t)$, $\xi\ll v_0 \tau_0$.
 The integral with respect to $\tau'$ is from $0$ to $t_{\rm enc}$, and we thus obtain from Eq.~(\ref{eqab})
 \begin{eqnarray}
\label{eqac}
&&\hspace*{-2.1em}\left\langle\exp\left(\frac{i}{\hbar}\Delta S_\gamma\right) \right\rangle\\&=& \exp\left(-\frac{2\Gamma v_0}{\xi\sqrt{\pi}}\left\langle\int_0^{t_{\rm enc}}\!\!\!\! d\tau'\int_{-\infty}^\infty \!\!\!d\tau\exp\left(-\frac{v_0^2\tau^2}{\xi^2} \right)  \right.\right.\nonumber\\&&\left.\left.\times \exp\left(-\frac{\left( t_l+2t_{\rm enc}-2\tau'\right)^2 }{\tau_0^2} \right)\right\rangle \right)\exp\left(-\Gamma t\right) . \nonumber 
\end{eqnarray}
We perform the two time integrals taking into account that only terms linear in $t_{\rm enc}$ will give a contribution when performing the $s, u$-integrals, see the remark after Eq.\ (\ref{eq22}). This finally yields for the action difference due to the perturbation
\begin{eqnarray}
&&\hspace*{-2.1em}\left\langle\exp\left(\frac{i}{\hbar}\Delta S_\gamma\right) \right\rangle \\&=& \left[1-2\Gamma\exp\left(-\frac{t_l^2}{\tau_0^2} \right)\left(t_{\rm enc}+\mathcal{O}\left(t_{\rm enc}\right) \right)   \right] \exp\left(-\Gamma t\right). \nonumber
\end{eqnarray}
The first term in the squared bracket gives a contribution independent of the perturbation and is canceled by the contribution coming from case (B).
We insert the remaining term into our expression for $m^{q_1}_{GOE}(t;\Gamma,\tau_0)$, evaluated as in
Sec.~\ref{fidelity-amplitude}, but that now depends additionally on the correlation time $\tau_0$ of the
explicitly time dependent part of the potential:
\begin{eqnarray}
m^{q_1}_{GOE}(t;\Gamma,\tau_0)\!\!&=&\!\!\int_{-c}^c\!\!\!\!dsdu\!\!\!\int_0^{t-2t_{\rm enc}}\!\!\!\!\!\!\!dt_l\frac{\left(t-2t_{\rm enc}-t_l \right)}{\Omega} \nonumber\\ &&\times \left( -2\Gamma\right) e^{\frac{i}{\hbar}su}\!\exp\!\!\left(-\Gamma t-\frac{t_l^2}{\tau_0^2}\right).\nonumber\\
\end{eqnarray}
After performing the remaining integrals we finally obtain the leading quantum correction in the presence of an explicitly time dependent perturbation
\begin{eqnarray}
\label{eq-time-dep}
m^{q_1}_{GOE}(t;\Gamma,\tau_0)\!\!\!&=&\!\!\!-\frac{2\Gamma}{T_{\rm H}}\left[\frac{\sqrt{\pi}}{2}\tau_0t{\rm Erf}\left(\frac{t}{\tau_0} \right)-\frac{\tau_0^2}{2}\right.\nonumber\\ &&\!\!\!\!\!\!\!\!\! \left.\times\left(1-\exp\left(-\frac{t^2}{\tau_0^2} \right)\right)  \right]\exp\left(-\Gamma t\right) , \nonumber\\
\end{eqnarray}
with the error function ${\rm Erf}(x)=\frac{2}{\sqrt{\pi}}\int_0^x dt\exp\left(-t^2 \right)$. 
By expanding ${\rm Erf}(x)$ we obtain in the limit $t \ll \tau_0$:
\begin{equation}
\label{eq-time-dep-2}
m^{q_1}_{GOE}(t\ll\tau_0;\Gamma) \simeq -\frac{\Gamma t^2}{T_{\rm H}} e^{-\Gamma t} 
\left[1 - \frac{1}{6}\left(\frac{t}{\tau_0}\right)^2 \right] \, ,
\end{equation}
indicating a small reduction of the interference term (\ref{eq19}),
$-(\Gamma t^2 / T_{\rm H})\exp(-\Gamma t)$ of the static case.

Much more interesting and relevant is the opposite limit, $t \gg \tau_0$, where we find
\begin{equation}
\label{eq-time-dep-3}
m^{q_1}_{GOE}(t\gg \tau_0;\Gamma) \simeq -\frac{\Gamma t^2}{T_{\rm H}} e^{-\Gamma t} 
\left[\sqrt{\pi} \frac{\tau_0}{t} - \left(\frac{\tau_0}{t}\right)^2 \right] \, .
\end{equation}
Quantum fidelity corrections  (\ref{eq19}) in the static case arise at time scales $t$ given by the geometrical
mean, $(\tilde{t}_\xi  T_{\rm H})^{1/2}$, of the Heisenberg time and the decay time 
$\tilde{t}_\xi\! =\! 1/2\Gamma$, Eq.~(\ref{decay-time-xi}). Equation (\ref{eq-time-dep-3})
hence implies, in view of the hierarchy of timescales, $\xi/v_0 \ll 2\tilde{t}_\xi < T_{\rm H}$, that 
the quantum fidelity contributions are suppressed (by $\tau_0 / t$) compared to the static case, 
if $\tau_0$ is much smaller than the above mentioned geometrical mean that usually represents
a large time scale. Together with the initial assumption, $\xi/v_0 \ll \tau_0$, we
can conclude that quantum fidelity contributions are suppressed for time-varying perturbations
with $\xi/v_0 \ll \tau_0 \ll (\tilde{t}_\xi T_{\rm H})^{1/2}$. 
Furthermore, such a suppression of this negative quantum correction implies that
upon reducing $\tau_0$, that is introducing faster time variations, the overall fidelity amplitude 
{\em increases} in the FGR regime and approaches $\exp(-\Gamma t)$.


\section{Fidelity }
\label{fidelity}

In this section we study the effect of the loop corrections on the semiclassical expression of the fidelity
itself, i.e.\ the average of the squared modulus of the fidelity amplitude: 
$M_{\rm sc}\left(t\right) =  \langle |m_{\rm sc}(t)|^2 \rangle $. We consider both the FGR and Lyapunov regime. 

\subsection{Fermi-golden-rule regime} 

In Ref.~\cite{gorin-2004} the fidelity has been addressed in the RMT approach within the
linear response approximation valid in the transition region between the perturbative and FGR regime.
Our semiclassical approach is not limited to such weak perturbations.
>From Eq.\ (\ref{eq1}) we obtain for the fidelity the semiclassical expression 
%
\begin{widetext}
\begin{eqnarray}
\label{eq40}
M_{\rm sc}(t)&=&\left(\dfrac{1}{2\pi\hbar} \right)^4 
\left\langle \int d\textbf{r}d{\textbf{r}}_0 d{\textbf
q} d\overline{\textbf{r}}d\overline{{\textbf{r}}}_0 d\overline{{\textbf{q}}}\,
\Psi_0^*\left({\textbf r}_0+\frac{\textbf{q}}{2}\right) \Psi_0\left({\textbf r}_0-\frac{\textbf{q}}{2}\right)
\Psi_0\left(\overline{{\textbf r}}_0+\frac{\overline{\textbf{q}}}{2}\right) 
 \Psi_0^*\left(\overline{{\textbf r}}_0-\frac{\overline{\textbf{q}}}{2}\right) \right.
\nonumber\\&& \times
\sum_{\gamma_1\left(\textbf{r}_0\rightarrow\textbf{r},t\right),\atop\gamma'_1\left(\textbf{r}_0\rightarrow\textbf{r},t\right)}
\sum_{\gamma_2\left(\overline{\textbf{r}}_0\rightarrow\overline{\textbf{r}},t\right),\atop\gamma'_2\left(\overline{\textbf{r}}_0\rightarrow\overline{\textbf{r}},t\right)}C_{\gamma_1}^{1/2}C_{\gamma'_1}^{*1/2} C_{\gamma_2}^{*1/2} C_{\gamma'_2}^{1/2} \\ 
&&\left. \times \exp \left[ {\frac{i}{\hbar} \left(
S_{\gamma_1}-S_{\gamma'_1}- S_{\gamma_2}+S_{\gamma'_2}-\frac{\textbf{q}}{2}\left(\textbf{p}_0^{\gamma_1}+
\textbf{p}_0^{\gamma'_1}\right)+\frac{\overline{\textbf{q}}}{2}\left(\textbf{p}_0^{\gamma_2}+
\textbf{p}_0^{\gamma'_2}\right)+\Delta S_{\gamma_1}-\Delta S_{\gamma_2} \right)  } \right] \right\rangle . \nonumber
\end{eqnarray}
\end{widetext}
As the fidelity decay in the FGR regime arises predominantly from uncorrelated trajectories $\gamma_1$ and $\gamma_2$ 
we can perform an disorder average of the phase differences $\Delta S_{\gamma_1}$ and $\Delta S_{\gamma_2}$ independently 
and obtain an expression containing four propagators 
that resembles the one obtained for the variance of the survival probability in Ref.~\cite{Gut},
with the only difference that the openness of the system is replaced by the presence of the random perturbation,
i.e.\  Eq.~(\ref{eq15}) by Eq.~(\ref{eq14}). In  Ref.~\cite{Gut} it could be shown that the variance vanishes in the
semiclassical limit $\hbar\rightarrow 0$. This argument still holds true here. 
Hence we conclude that there remain also only negligible corrections to the fidelity apart from those arising from
the squared modulus of the average fidelity amplitude already calculated, implying 
$M_{\rm sc}(t) = \langle \left|m_{\rm sc}(t)\right|^2 \rangle\approx  \left\langle \left|m_{\rm sc}(t)\right|\right\rangle^2$. Hence we can simply use the right hand side of the last relation to obtain, 
to leading order in $t/T_{\rm H}$, the following for the GOE case
\begin{equation}
M^{q_1}_{\rm GOE}(t) = \exp(-2\Gamma t) \left(1 - 2\frac{\Gamma t^2}{T_{\rm H}} \right) \, , 
\label{Fidelity-GOE}
\end{equation}
and for the GUE case
\begin{equation}
M^{q}_{\rm GUE}(t) = \exp(-2\Gamma t) \left(1 - \frac{2\Gamma t^3}{3T_{\rm H}^2} \right) \, .
\label{Fidelity-GUE}
\end{equation}

\subsection{Lyapunov regime}
\label{lyapunov}

Above we considered the case where the disorder average in the calculation of $M_{\rm sc}(t)$ can be performed
independently for the trajectory pairs occurring in the calculation of each $m_{\rm sc}(t)$. However, 
already on the level of the diagonal approximation there is a further contribution originating from configurations 
where all four trajectories are too close together to perform the disorder average independently.
The regime of large $\Gamma$ where the FGR terms (\ref{Fidelity-GOE},\ref{Fidelity-GUE}) 
are rapidly decaying and this contribution becomes important is referred to as the Lyapunov regime, because it
decays as ${\rm e}^{-\lambda t}$ \cite{Jalabert}.

Here we examine whether additional contributions may arise from loop diagrams. To this end
we briefly review the semiclassical calculation of the diagonal contribution to the fidelity 
in the Lyapunov regime \cite{Jalabert} 
and consider afterwards the role of trajectories differing at encounters.

Starting from Eq.\ (\ref{eq40}). 
one performs the disorder average along the trajectories no longer independently for both action differences 
$\Delta S_{\gamma_1}$ and $\Delta S_{\gamma_2}$. For two nearby trajectories ${\gamma_1}, {\gamma_2}$, one instead 
linearizes the motion of one trajectory around the other to obtain
\begin{equation}
\label{eq100}
\Delta S_{\gamma_1}-\Delta S_{\gamma_2}=\int_0^t dt' {\pmb{\nabla}}L^\Sigma_{\gamma_1}\left(t'\right) \left[
\textbf{q}_{\gamma_1}\left(t'\right)-\textbf{q}_{\gamma_2}\left(t'\right)\right] \, , 
\end{equation}
where  $L^\Sigma_{\gamma_1}\left(t'\right)$  is the Lagrangian associated with the disorder as in Eq.\ (\ref{eqa}),
and  $\textbf{q}_{\gamma_i}\left( t'\right) $ denotes the coordinates of the trajectories $\gamma_i$  at time $t'$.
The difference $\textbf{q}_{\gamma_1}\left(t'\right)-\textbf{q}_{\gamma_2}\left(t'\right)$ can then be expressed by 
the difference of the final points $\textbf{r}$ and $\overline{\textbf{r}}$ of $\gamma_1$ and $\gamma_2$, respectively,
using the exponential separation of long neighboring trajectories in the chaotic case.
Assuming again a Gaussian distribution of the random variable, $\Delta S_{\gamma_1}-\Delta S_{\gamma_2}$, as
in the FGR regime for $\Delta S_{\gamma}$ in Eq.\ (\ref{eqe}), one obtains 
\begin{eqnarray}
\label{eq102}
&&\hspace*{-1.5em}\left\langle \exp\left[ \frac{i}{\hbar}\left( \Delta S_{\gamma_1}-\Delta S_{\gamma_2}\right) \right] 
\right\rangle \\  
&=& \exp\left[-\frac{1}{2\hbar^2} \int_0^t dt'\!\! \int_0^t dt''e^{\lambda\left(t'+t''-2t \right)} 
C_{\pmb{\nabla}\Sigma}\left(\textbf{r}_{1}-\textbf{r}_{2} \right)^2\right]\nonumber
\end{eqnarray}
with the force correlator $C_{\pmb{\nabla}\Sigma}   \equiv\left\langle {\pmb{\nabla}L^\Sigma_{\gamma_1}\left(t'\right) }{\pmb{\nabla}L^\Sigma_{\gamma_2}\left(t''\right)}\right\rangle$.

Replacing in Eq.~(\ref{eq102}) the  second time integral by one with respect to $t'-t''$ and taking into account the 
short range behavior of the force correlator  (it vanishes on scales larger than the correlation length $\xi$)
one can perform the $(t'-t'')$-integral in the range from $-\infty$ to $\infty$. Further evaluation of the $t'$-integral, 
where one neglects contributions from the lower limit because they are damped by a factor ${\rm e}^{-\lambda t}$, 
eventually gives 
\begin{equation}
\label{eq104}
\left\langle \exp\left[ \frac{i}{\hbar}\left( \Delta S_{\gamma_1}-\Delta S_{\gamma_2}\right) \right] \right\rangle
=\exp\left[-\frac{A}{2\hbar^2}\left|\textbf{r}_1-\textbf{r}_2\right|^2\right]  \, .
\end{equation}
Here the constant $A$ depends  on $\xi$ and the disorder strength. 
Equation~(\ref{eq104}) is afterwards inserted into the full expression for the fidelity 
finally yielding a contribution proportional to ${\rm e}^{-\lambda t}$ within the diagonal approximation \cite{Jalabert}.
We note that when performing the two time integrals in Eq.~(\ref{eq102}), only contributions 
from $t'$ and $t''$ close to $t$ mattered.

 
We now consider a possible effect of non-diagonal loop contributions. The basic contribution of this kind would originate 
from trajectories $\gamma_1$ and $\gamma_1'$  forming pairs as depicted in Fig.\ 1 and two nearly identical trajectories 
$\gamma_2$ and $\gamma_2'$. 
When calculating 
$C_{{\pmb\nabla}\Sigma}$, correlations between points of $\gamma_1$ traversed at different times could get important 
as the latter quantity depends on the difference of the positions along $\gamma_1$ at two different times. 
However, as we noted below Eq.\ (\ref{eq104}), correlations in the Lyapunov regime only matter between
points at the end of the orbit, i.e.\ between points, where the motion can still be approximated by a free motion. 
Compared to that distance, an encounter stretch is exceedingly long, so that correlations between \textit{different} 
encounter stretches cannot play a role for this contribution. This implies that there is no significant effect of 
such orbit pairs in the Lyapunov regime, because there exists, apart from the action difference and the weight 
function, no further phase change induced by the encounter (depending on the $s,u$-coordinates). 

The same reasoning can be directly carried over to correlated orbits differing by an arbitrary number of encounters,
and hence no off-diagonal interference contributions from orbits differing in encounters can be obtained in the 
Lyapunov regime. This constitutes another distinct difference between FGR- and Lyapunov decay.

\section{Conclusions and Outlook}

In this work, we have presented a semiclassical theory for the time decay of the quantum fidelity
amplitude and the fidelity, or Loschmidt echo, for quantum systems that are chaotic in
the classical limit. Our focus has been on the calculation of quantum contributions to the fidelity
amplitude beyond the leading term, $\sim e^{-\Gamma t}$, which we showed to arise semiclassically
from interference between pairs of topologically related trajectories involving classical
correlations between paths. Applying advanced semiclassical methods to deal with multiple pairs
of correlated orbits, we derived quantum corrections which agree with supersymmetric results in
the universal Fermi golden rule regime. Deriving recursion relations for the semiclassical objects, 
we could show that the semiclassical formulation of the fidelity amplitude
obeys unitarity and we could confirm the interesting relation~(\ref{relationship}) 
between the fidelity amplitude and the parametric form factor to any order in $t/ T_{\rm H}$ for times
smaller than the Heisenberg time $T_{\rm H}$.

Besides, our semiclassical approach provides insight into the interference mechanisms underlying
quantum fidelity decay. The leading quantum corrections in a $t/T_{\rm H}$-expansion, Eq.~(\ref{eq19}), can
be regarded as arising from a weak-localization type effect leading to a reduction in the fidelity
amplitude in the time-reversal case, which is susceptible to time-reversal symmetry breaking,
e.g.\ through a magnetic field. Moreover, the present approach enables an interpretation of the 
fidelity dependence on the characteristics of the assumed random perturbation, in particular the
spatial correlation length $\xi$ and the time correlation $\tau_0$. The first appears as a
further length scale in the Ehrenfest time dependence of the fidelity amplitude. 
Even more interestingly, through $\tau_0$ we can estimate effects from time- or 
frequency-dependent perturbing sources, such
as phonons in solids. As a result, we could derive explicit expressions, see
Eqs.~(\ref{eq-time-dep},\ref{eq-time-dep-2},\ref{eq-time-dep-3}), 
showing how the negative weak localization-type contribution is suppressed
with decreasing $\tau_0$, and thereby the fidelity is counter-intuitively enhanced.

One main open question arising from this work is how to extend the present
semiclassical formalism which is limited to time scales below the Heisenberg time, to longer
times. This would also open up a way towards a semiclassical understanding of the fidelity
amplitude revival  close to $T_{\rm H}$ in the FGR regime \cite{Sto-2004,Sto-2005}. Moreover,
semiclassics beyond $T_{\rm H}$ would allow for extending our analysis to a further decay regime,
namely for treating the perturbative regime of very weak perturbations corresponding to
energy scales below the mean level spacing.

Throughout this work we considered fidelity decay due to a global perturbation affecting the whole 
configuration space or system boundary. Recently, the complementary case of local, not necessarily
weak, perturbations has been assessed semiclassically on the level of the diagonal approximation
\cite{Gou-07,Gou-08}. There, instead of the FGR- and Lyapunov decay, corresponding decay regimes, 
namely a FGR-type and a new `escape rate' regime, have been identified exhibiting in particular a 
peculiar non-monotonic crossover between the two regimes when passing from weak to strong local perturbation. 
Quantum corrections in both regimes would originate from the same diagrams as for semiclassical decay of
open quantum systems, and a semiclassical treatment of the latter \cite{Wal,Gut} can be readily applied
to obtain a more complete picture of fidelity decay for local perturbations.
Again, accessing time scales beyond the Heisenberg time remains there as a challenge for the future.

Finally, our semiclassical techniques may be applied to extend a recent semiclassical analysis
\cite{CJP08PRL} of the mechanism of time-reversal mirroring \cite{Fink}, a concept related to the 
fidelity, by including quantum interference effects from correlated mirror trajectories.

\section{Acknowledgments}

We thank \.Inan\c{c} Adagideli, Peter Braun, Arseni Goussev, Philippe Jacquod, Rodolfo Jalabert, Heiner Kohler, Cyril Petitjean, Thomas Seligman and
Hans-J\"urgen St\"ockmann for useful discussions and Arseni Goussev, Rodolfo Jalabert for a critical reading of the manuscript.
Financial support by the Deutsche Forschungsgemeinschaft within SFB/TR 12 (BG), 
GRK 638 (DW, MG, JK, KR), FOR 760 (KR) and from the Alexander von Humboldt Foundation (JK) is 
gratefully acknowledged.

\end{document}